\documentclass[twocolumn,times]{aastex63}
\usepackage{amsmath}
\usepackage{color}

\newcommand{\Galex}{\textit{GALEX}}
\newcommand{\Swift}{\textit{Swift}}
\newcommand{\Chandra}{\textit{Chandra}}
\newcommand{\XMM}{\textit{XMM-Newton}}
\newcommand{\Herschel}{\textit{Herschel}}

\defcitealias{Hodges-Kluck2016}{HK16}

\begin{document}

\shorttitle{EXTENSIVE WIND CONE IN NGC 3079}
\shortauthors{HODGES-KLUCK ET AL.}

\title{A 60~kpc Galactic Wind Cone in NGC~3079}

\author{Edmund J. Hodges-Kluck}
%\affil{Department of Astronomy, University of Maryland, College Park, MD 20740, USA}
\affil{Code 662, NASA Goddard Space Flight Center, Greenbelt, MD 20771, USA}
\email{edmund.hodges-kluck@nasa.gov}

\author{Mihoko Yukita}
\affil{Code 662, NASA Goddard Space Flight Center, Greenbelt, MD 20771, USA}
\affil{Henry A. Rowland Department of Physics and Astronomy, Johns Hopkins University, Baltimore, MD 21218, USA}

\author{Ryan Tanner}
\affil{Code 662, NASA Goddard Space Flight Center, Greenbelt, MD 20771, USA}
\affil{Universities Space Research Association, 7178 Columbia Gateway Dr, Columbia, MD 21046}

\author{Andrew F. Ptak}
\affil{Code 662, NASA Goddard Space Flight Center, Greenbelt, MD 20771, USA}

\author{Joel N. Bregman}
\affil{Department of Astronomy, University of Michigan, Ann Arbor, MI 48109, USA}

\author{Jiang-tao Li}
\affil{Department of Astronomy, University of Michigan, Ann Arbor, MI 48109, USA}

\begin{abstract}
Galactic winds are associated with intense star formation and AGNs. Depending on their formation mechanism and velocity they may remove a significant fraction of gas from their host galaxies, thus suppressing star formation, enriching the intergalactic medium, and shaping the circumgalactic gas. However, the long-term evolution of these winds remains mostly unknown. We report the detection of a wind from NGC~3079 to at least 60~kpc from the galaxy. We detect the wind in FUV line emission to 60~kpc (as inferred from the broad FUV filter in \Galex) and in X-rays to at least 30~kpc. The morphology, luminosities, temperatures, and densities indicate that the emission comes from shocked material, and the O/Fe ratio implies that the X-ray emitting gas is enriched by Type~II supernovae. If so, the speed inferred from simple shock models is about 500~km~s$^{-1}$, which is sufficient to escape the galaxy. However, the inferred kinetic energy in the wind from visible components is substantially smaller than canonical hot superwind models. 
\end{abstract}

\keywords{Galactic Winds -- Starburst Galaxies -- Late-type Galaxies}

\section{Introduction}
\label{section.intro}

Intense periods of star formation can lead to multi-phase, galaxy-scale outflows driven by the energy and momentum of stellar winds and supernovae. These galactic winds can have speeds of hundreds to thousands of km~s$^{-1}$ and are the most extreme instance of stellar feedback \citep{Veilleux2005,Heckman2017,Rupke2018}. Winds have the potential to regulate or even quench star formation in their host galaxies, and are perhaps the main source of metals in the intergalactic medium. It is generally accepted that star formation and AGN are the source of galactic winds, but how these determine the morphology, kinematics, composition, and how much (if any) material completely escapes the galaxy are unsettled questions. Plausible theoretical models vary widely in their answers to these questions \citep[for a recent theoretical review, see][]{Zhang2018}. 

Addressing these uncertainties requires measuring the evolution of wind properties with distance from the galaxy, such as the temperature, density, and velocity phase diagrams. This is challenging because winds expand and decrease in surface brightness, so beyond a few kpc from the galaxy winds have  primarily been studied in absorption. For example, \citet{Kacprzak2012} found using the MAGIICAT dataset of Mg~{\sc ii} absorbers that absorption occurs most frequently within $\sim 20^{\circ}$ of the minor axis (outflows) and major axis (accretion) of the host galaxies \citep[although this is not generally true beyond 40~kpc;][]{Bordoloi2014}, while \citet{Heckman2017b} showed with the COS-Burst survey that Ly$\alpha$ absorbers around starburst galaxies have roughly twice the virial velocity, suggesting rapid outflows.  

Nevertheless, searching for emission at large radii is important because it enables detailed study of winds from individual objects, whereas pencil-beam absorption studies must build up samples of absorption systems around similar galaxies or focus on well-placed sightlines. This is limiting because the wind properties are sensitive to the characteristics of the generative starburst or AGN, which may differ substantially even in similar galaxies. In addition to emission, scattered Lyman-$\alpha$ can probe winds  in individual galaxies \citep[e.g.,][]{Duval2016}, but as it does not directly probe ionized gas and can have complex line-of-sight radiative transfer even after scattering it is not a primary tool. One of the best ways to search for emission from winds, which is expected to be primarily line emission, is to use integral field spectrographs on large telescopes \citep[e.g.,][]{Finley2017}. Recently, \citep{Rupke2019} reported the discovery of a 100~kpc wind in SDSS~J211824.06+001729.4 (``Makani'') at $z=0.459$ using the integral field unit on the Keck observatory. However, it remains worthwhile to search for extended emission from nearby winds because they can be observed with current instruments in nearly every waveband. 

Here we present evidence for emission from shocked gas in a biconical outflow extending at least 60~kpc from NGC~3079, using far ultraviolet (FUV) and X-ray data. NGC~3079 is a well known starburst and AGN galaxy with an extensive radio halo \citep{Irwin2003}, a nuclear H$\alpha$ and X-ray bubble \citep{Cecil2001}, and diffuse nuclear hard X-rays \citep{JiangtaoLi2019}. A biconical outflow at lower latitudes has already been established from previous H$\alpha$, X-ray, and UV studies \citep[e.g.,][]{Fabbiano1992,Strickland2004}. There is also a suggestion that the extensive \ion{H}{1} tail seen behind the companion NGC~3073 is due to stripping by the wind from NGC~3079 \citep{Irwin1987}, as the ambient halo density required to explain it via stripping in the hot halo of NGC~3079 or the tidal forces needed to strip the gas are inconsistent with existing data \citep{Shafi2015}. If so, this would indicate a wind extended significantly beyond the radio halo, although the prospect of NGC~3073 itself driving an outflow has not been thoroughly explored. The starburst in NGC~3079 is fueled by $3\times 10^8 M_{\odot}$ of molecular gas in the nucleus \citep{Sofue2001} with a nuclear star-formation rate (SFR) of $2.6 M_{\odot}$~yr$^{-1}$ \citep{Yamagishi2010}. 

In the following sections we describe the data sources (Section~\ref{section.data}), the search for extended diffuse emission and structure (Section~\ref{section.filaments}), and our interpretation of this structure based on the X-ray properties (Sections~\ref{section.xrays} and \ref{section.interpretation}). We close with a summary in Section~\ref{section.summary}.  We adopt a distance of $d=19$~Mpc for NGC~3079 \citep{Springob2009}, which corresponds to a scale of 5.5~kpc~arcmin$^{-1}$ and is the median redshift-independent distance from the NASA/IPAC Extragalactic Database\footnote{http://ned.ipac.caltech.edu}.

\section{Observations and Data}
\label{section.data}

\begin{figure*}[htp]
    \centering
    \includegraphics[width=0.95\textwidth]{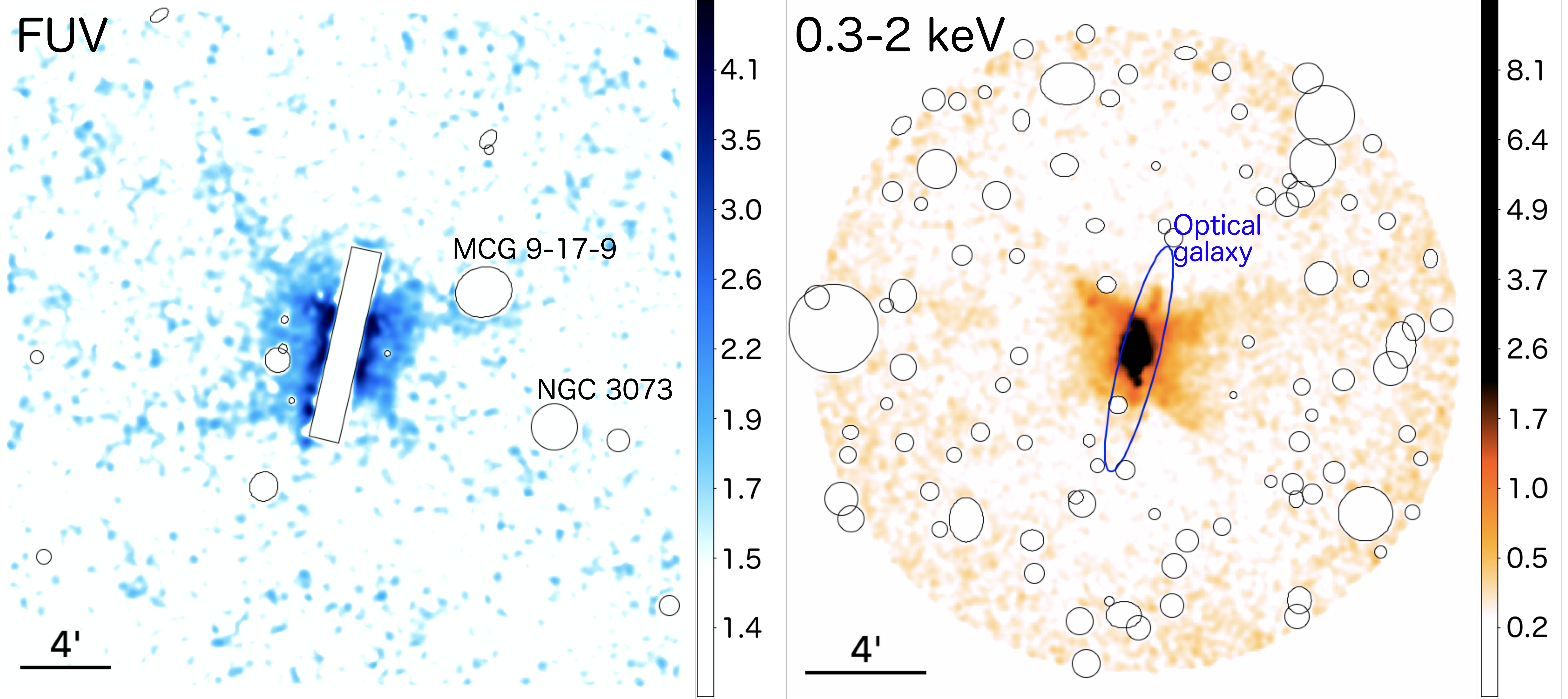}
    \caption{\textit{Left}: \Galex\ FUV image of NGC~3079, corrected for the galaxy light scattered into the wings of the PSF (see text for an explanation). The galaxy itself is clipped out near $R_{25}$ and bright source masks are indicated in black (many more small point sources were excised). Several filament candidates form an ``X'' shape. The color table is in units of $10^{-4}$~counts~s$^{-1}$. \textit{Right}: 0.3-2~keV \XMM\ image of the same field with source masks shown in black and the optical galaxy ($R_{25}$) shown as a blue ellipse. The same ``X'' shape seen in previous \Chandra\ images is evident. The color table is in units of $10^{-5}$~counts~s$^{-1}$. Both images have been smoothed with a Gaussian kernel and clipped at the mean background. }
    \label{fig:images}
\end{figure*}

The primary data sets used in this paper are a 16.1~ks \Galex\ (Galaxy Evolution Explorer) observation of NGC~3079 and 300~ks of new and archival \XMM\ observations. 231~ks of new observations were obtained under \XMM\ project 080271 (PI: Hodges-Kluck). We also used archival \Chandra, \textit{Neil Gehrels Swift Observatory} (\Swift), \textit{Infrared Astronomy Satellite} (IRAS), and \Herschel\ observations as complementary data sets. 

% Telescope | Instrument/Filter | Date | ObsID | Exposure | GTI
\begin{deluxetable}{lccccc}
\tablenum{1}
\tabletypesize{\scriptsize}
\tablecaption{\label{table.obs} UV and X-ray Observations in this Paper}
\tablewidth{0pt}
\tablehead{
\colhead{(1)} & \colhead{(2)} & \colhead{(3)} & \colhead{(4)} & \colhead{(5)} & \colhead{(6)} \\
\colhead{Telescope} & \colhead{Instrument} & \colhead{Date} & \colhead{ObsID} & \colhead{Exposure} & \colhead{GTI} \\ 
 &  &  &  & \colhead{($10^3$~s)} & \colhead{($10^3$~s)}
}
\startdata
GALEX   & FUV    & 2005-02-24 & NGA\_NGC3079 & 16.1 & 16.1 \\ 
GALEX   & NUV    & 2005-02-24 & NGA\_NGC3079 & 16.1 & 16.1 \\ 
Swift   & UVW1   & 2013-11-12 & 80030001     & 0.3  & 0.3  \\
Swift   & UVM2   & 2008-02-26 & 37245001     & 8.5  & 8.5  \\
Swift   & UVW2   & 2009-02-27 & 37245002     & 1.2  & 1.2  \\
Swift   & UVW2   & 2013-11-12 & 80030001     & 0.6  & 0.6  \\
Swift   & UVW2   & 2014-04-04 & 91912001     & 0.7  & 0.7  \\
Swift   & UVW2   & 2014-04-06 & 91912002     & 4.9  & 4.9  \\
XMM	    & EPIC   & 2001-04-13 & 110930201    & 25.3 & 3.1  \\
XMM	    & EPIC   & 2003-10-14 & 147760101    & 44.4 & 6.6  \\
XMM	    & EPIC   & 2017-11-01 & 802710101    & 22.8 & 17.0 \\
XMM	    & EPIC   & 2017-11-03 & 802710201    & 22.4 & 7.1  \\
XMM	    & EPIC   & 2017-11-05 & 802710301    & 22.4 & 9.1  \\
XMM	    & EPIC   & 2017-11-09 & 802710401    & 22.4 & 3.8  \\
XMM	    & EPIC   & 2017-11-15 & 802710501    & 22.4 & 11.3 \\
XMM	    & EPIC   & 2017-11-23 & 802710601    & 22.4 & 14.1 \\
XMM	    & EPIC   & 2017-11-27 & 802710701    & 22.4 & 12.5 \\
XMM	    & EPIC   & 2018-04-17 & 802710801    & 26.0 & 22.0 \\
XMM	    & EPIC   & 2018-04-21 & 802710901    & 22.4 & 18.6 \\
XMM	    & EPIC   & 2018-04-23 & 802711001    & 25.3 & 21.0 \\
Chandra & ACIS-S & 2001-03-07 & 2038         & 26.6 & 26.6 \\
Chandra & ACIS-S & 2018-01-30 & 19307        & 53.2 & 53.2 \\
Chandra & ACIS-S & 2018-02-01 & 20947        & 44.4 & 44.4 \\
\enddata
\end{deluxetable}

\subsection{GALEX}
We used the FUV and NUV pipeline-processed \Galex\ images. The FUV filter has an effective wavelength of 1542\AA\ and a width (FWHM) of 228\AA. The NUV filter has an effective wavelength of 2274\AA\ and a width of 796\AA. The on-axis angular resolution of the 50~cm telescope is about 4.2\arcsec\ in the FUV and 5.3\arcsec\ in the NUV channel. To prepare the images for analysis, we clipped the image to a 25\arcmin$\times$25\arcmin\ square around the galaxy, identified and masked point sources at least 3$\sigma$ above background, and masked extended sources through visual inspection. The size of the region of interest is limited by the significantly increased noise and image reconstruction artifacts near the edge of the \Galex\ field, and no overlapping \Galex\ image is nearly as deep. Since we are interested in diffuse light near the bright galaxy, we corrected for galactic light scattered into the halo region by the wings of the point-spread function (PSF). We followed the procedure described in \citet{Hodges-Kluck2016} (HK16), which involves subtracting a scaled convolution of the galaxy image (i.e., within the optical $R_{25}$) with the PSF from the raw image. The scale factor depends on the PSF and accounts for the fact that the galaxy image has already been convolved with the PSF. We used a PSF constructed in \citetalias{Hodges-Kluck2016} that is 15\arcmin\ in radius. There are no other extended sources in the field bright enough to require this procedure. The resultant image is shown in Figure~\ref{fig:images}. 

\subsection{XMM}
We processed each \XMM\ dataset using Science Analysis Software (SAS) v17.0.0 and applied standard procedures to extract events files for each camera (two MOS and one pn). First, we used the {\tt emchain} and {\tt epchain} scripts to filter and calibrate events and produce analysis-ready event lists for the MOS and pn cameras (for the pn we also extracted out-of-time events). We then used the XMM Extended SAS (E-SAS) software \citep{Snowden2004} to further filter the data, including removing background flares (through {\tt mos\_filter} and {\tt pn\_filter}), masking point sources (through modified {\tt cheese} masks), and extracting spectra with the quiescent particle background and an estimate of the solar wind charge exchange subtracted (through {\tt mos-spectra}, {\tt mos\_back}, {\tt pn-spectra}, and {\tt pn\_back}). After filtering, about 150~ks of good time remained (Table~\ref{table.obs}). We used the background estimates to create filtered images from each detector, and combined these images across all exposures for the image analysis. The combined X-ray image is several times more sensitive to faint point source emission than any individual exposure and so we identified and masked additional sources with {\tt edetect\_chain} to highlight the diffuse emission. In general, the source masks extend to where the light from the wings of the PSF is well below the background rather than to a fixed encircled energy fraction. 

\begin{figure*}
    \centering
    \includegraphics[height=3in]{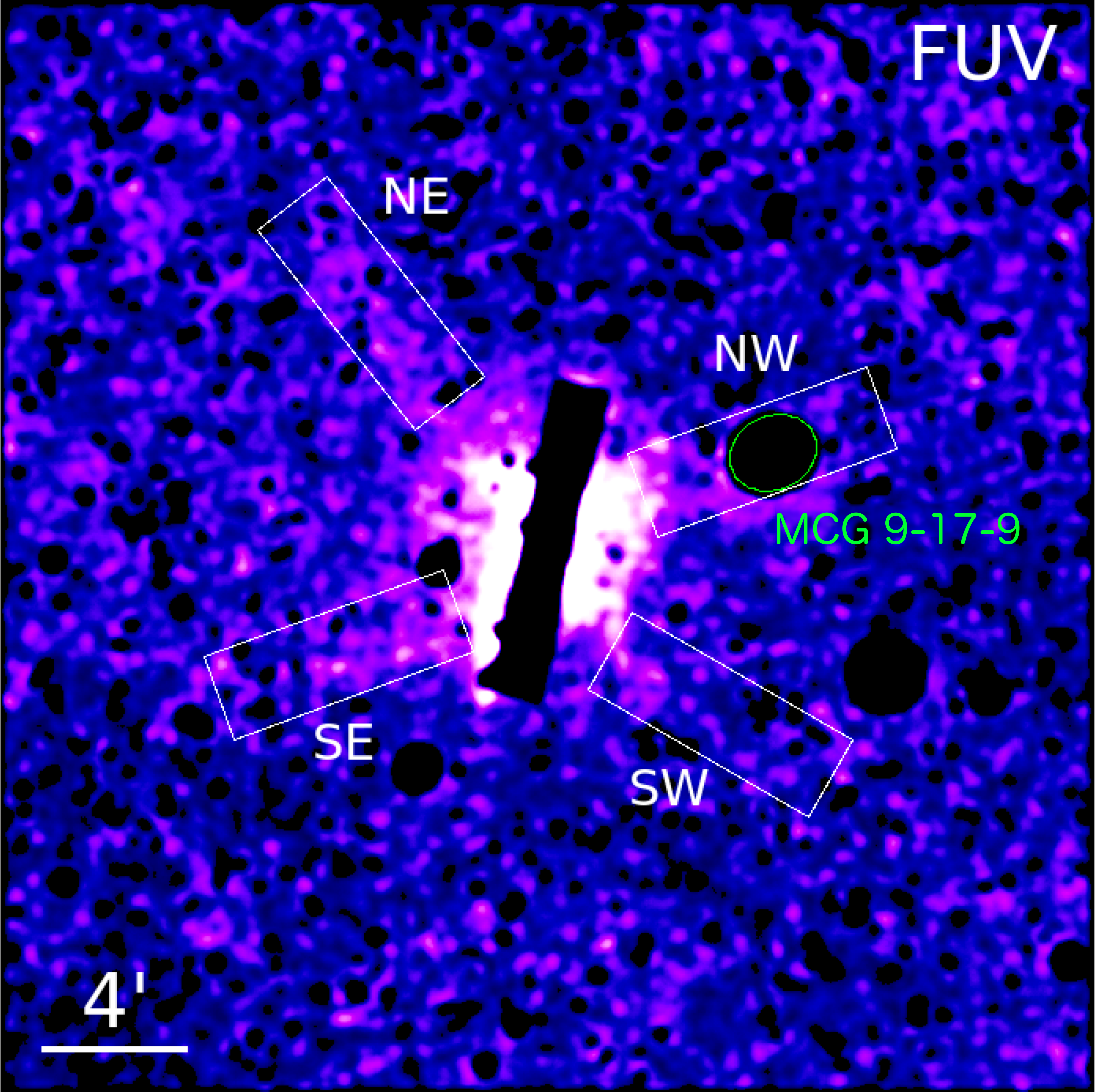}
    \includegraphics[height=3in]{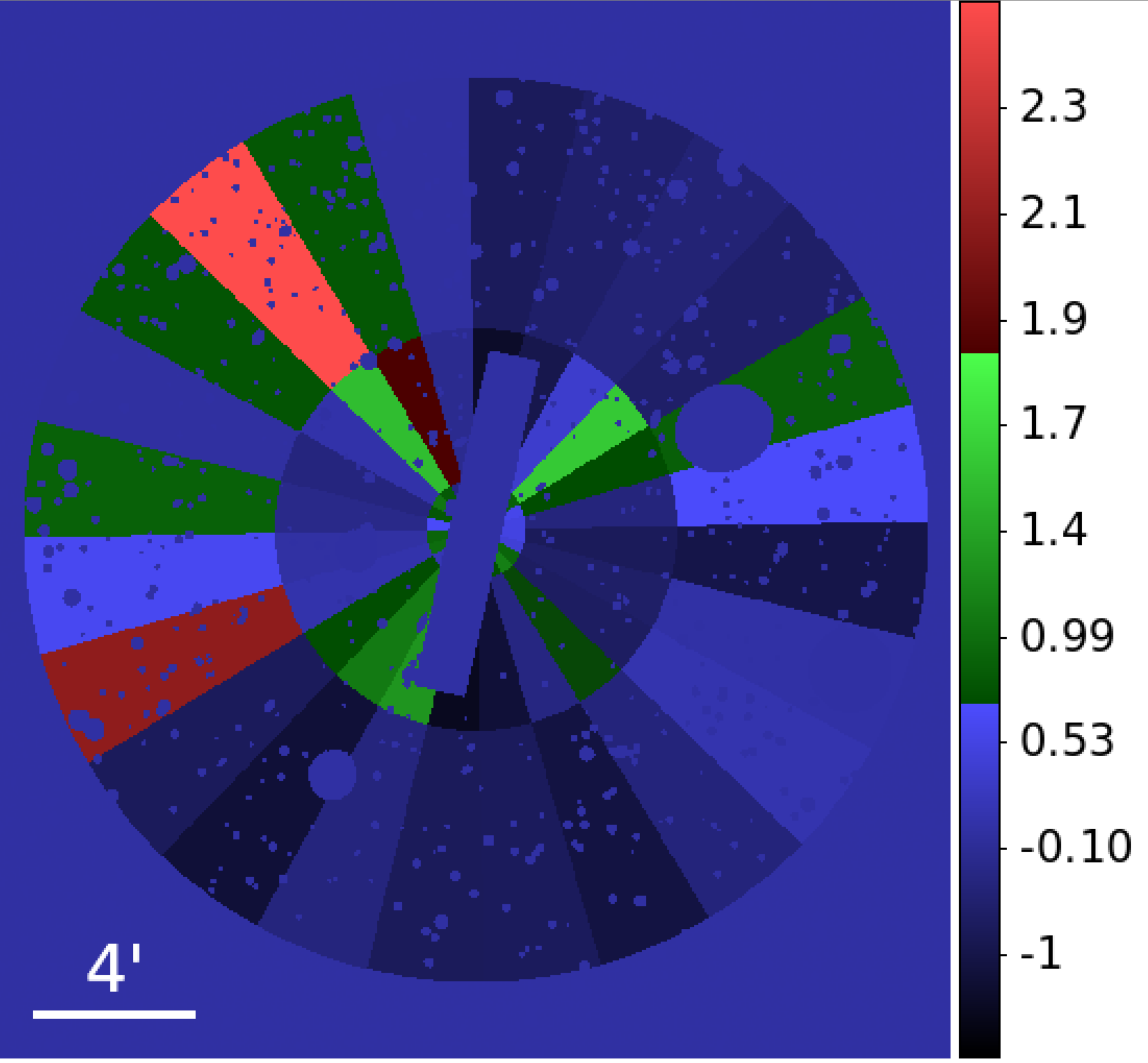}
    \caption{\textit{Left}: Four candidate filaments identified in the \Galex\ FUV image. The maximum apparent extent is about 66~kpc from the galaxy nucleus on the east side. The existence and extent of the northwest filament is unclear because it coincides with the companion galaxy, MCG 9-17-9, located 33~kpc from NGC~3079. 
    \textit{Right}: An azimuthal significance map between $R=$0--1.25~arcmin, 1.25--5.0~arcmin, and 5.0--11.25~arcmin. Each segment is 15$^{\circ}$ wide. In each ring the color represents $(X-\mu)/\sigma$, where $X$ is the mean value in each segment, $\mu$ is the mean for the ring, and $\sigma$ is the standard deviation of values in the ring. The colors have been biased towards positive deviations.  
    }
    \label{fig:fuv_filaments}
\end{figure*}

For the image analysis we adopt the $0.3-2$~keV bandpass to maximize the signal from hot gas. The combined $0.3-2$~keV image is shown in Figure~\ref{fig:images}. The sensitivity is not uniform across the field, and this image has been clipped to the region where the effective exposure exceeds about 80\% of the total. However, there is clearly a ring of increased noise that limits the search for extended diffuse emission. The primary astrophysical backgrounds are the hot gas in and around the Milky Way and solar wind charge exchange, whereas the instrumental backgrounds include residual soft proton flaring (below the flare detection threshold) and a strong Al~K$\alpha$ line (1.49~keV) from the detector. The astrophysical backgrounds are vignetted, the protons are centrally concentrated but not optically vignetted, and the detector background is not vignetted. This means that simple exposure correction can be misleading. We follow \citet{Anderson2014} in accounting for backgrounds (but not in fitting a radial profile) in the $0.3-2$~keV bandpass, which involves fitting the soft X-ray background spectrum at large radii to obtain its physical normalization, using the unexposed chip corners (through the ESAS software) to make non-vignetted background maps, and ``exposure correcting'' using a factor based on the vignetting function as a function of radius. 
\subsection{Auxiliary Data}
We used the \textit{uvw1} ($\lambda$2600\AA), \textit{uvm2} ($\lambda$2200\AA), and \textit{uvw2} ($\lambda$2000\AA) \Swift\ Ultraviolet and Optical Telescope (UVOT) images as a check on the deeper \Galex\ data. We previously processed these data in \citetalias{Hodges-Kluck2016}, including the removal of persistent, diffuse scattered light artifacts and correcting for the PSF scattering from the galaxy. Multiple short exposures were combined to form single images for each filter. 

The 124~ks of archival \Chandra\ Advanced CCD Imaging Spectrometer (ACIS) data was used to identify X-ray point sources to remove from the \XMM\ data and to search for diffuse emission. Both data sets were centered on the ACIS-S3 chip. Unfortunately, the longer \Chandra\ data sets were obtained after the detector lost most of its soft sensitivity ($E<1$~keV) due to molecular contaminant buildup on the detector window. Thus, the \Chandra\ sensitivity to hot gas is much worse than with \XMM. We reprocessed the data using the \Chandra\ Interactive Analysis of Observations (CIAO) v4.11 software\footnote{http://cxc.harvard.edu/ciao/}. We used the standard {\tt chandra\_repro} script to filter and grade the events. We then searched for background flares in the light curve for the whole ACIS-S3 chip and excluded periods where the count rate exceeded 3$\sigma$ above background but found no significant flaring. For the imaging analysis, we reprojected and combined the data sets using the {\tt merge\_obs} script. We then identified and removed point sources, and restricted the bandpass to $0.3-2$~keV.

We also used the 100~$\mu$m IRAS and pipeline-processed \textit{Herschel} data in several bands from 70-160~$\mu$m without further processing.

\section{An Extended Wind Cone}
\label{section.filaments}

We searched for extended emission in the \Galex\ and X-ray images (Figure~\ref{fig:images}). Both the FUV and X-ray images show the characteristic ``X'' shape of a biconical wind known to exist around NGC~3079 from previous H$\alpha$ and \Chandra\ observations, but the filaments along the wind edges (identified in Figure~\ref{fig:fuv_filaments}) appear to extend significantly farther than previously known, especially in the FUV image where the eastern filaments extend at least to 60~kpc from the galactic center. The western filaments appear more truncated but extend at least to 30-40~kpc from the nucleus. Meanwhile, the \XMM\ image shows more extended X-ray emission on the west side with a maximum extent of about 40~kpc, and to 30-35~kpc on the east. The emission may extend farther but the sensitivity declines significantly outside $R \sim 40$~kpc. 

\subsection{Filaments}

The filaments are resolved, with widths exceeding several kpc, although it is difficult to measure their widths owing to the low surface brightness. This also makes it difficult to objectively define filaments. Here we assumed that they simply follow the direction of the brighter emission seen at lower latitudes and defined rectangular boxes 38$\times$13~kpc wide that start at the edge of the brighter reflection nebula reported at lower latitudes in \citet{Hodges-Kluck2016}. 7~arcmin in length (38~kpc) appears well matched to the data, but was defined by eye. We measured the fluxes in these boxes, which are given in Table~\ref{table.filaments} and overlaid on Figure~\ref{fig:fuv_filaments}. 

Detecting low surface brightness emission requires accounting for true variation in the background across the field of view. This variation, rather than Poisson noise, limits the sensitivity. We defined multiple source-free regions around the galaxy at the outskirts of the clipped field of view and measured the mean background in each, as well as the standard deviation among fields. We find $B_{\text{NUV}} = (2.17\pm0.01) \times 10^{-3}$~counts~s$^{-1}$~pix$^{-1}$ and $B_{\text{FUV}} = (1.44\pm0.02) \times 10^{-4}$~counts~s$^{-1}$~pix$^{-1}$, which is consistent with typical GALEX values. The fluxes in each filament region were measured and converted to magnitudes using the AB zero points of 20.08 and 18.82~mag for the NUV and FUV, respectively. The sensitivities imply 3$\sigma$ upper limits for these boxes of around 20.4 and 21.0~mag for the NUV and FUV, respectively (surface brightness of 23.5 and 24.1~mag~arcmin$^{-2}$). 

Each of the FUV filaments is clearly detected (for a 3$\sigma$ threshold). The luminosities range from 2-5$\times 10^{40}$~erg~s$^{-1}$ and their total luminosity is about 1\% of the galaxy FUV luminosity. In contrast, no filaments are detected in the NUV image. The limits imply FUV/NUV luminosity ratios greater than 2 in the NE, NW, and SE filaments, corresponding to FUV$-$NUV colors of less than $0.4$~mag. The SW filament is the weakest detection in the FUV, but it is also not present in the NUV and Figure~\ref{fig:fuv_filaments} suggests that it may not extend the full length of the (uniform) 38~kpc box. 

Although there are no extended ($R>15$~kpc) NUV filaments, at lower latitude diffuse UV continuum emission is detected in the FUV, NUV, and the UVOT images. The morphology in each filter agrees well and the measured fluxes are similar. This emission was ascribed to scattering by circumgalactic dust in \citetalias{Hodges-Kluck2016}, indicating that the FUV filaments are something else. However, since we defined the filaments based on the presumption that they continue from spurs seen at lower latitude it is likely that not all of the UV light seen at lower latitude comes from dust. 

The measured fluxes show that there is some extended emission in the boxes that we defined, but this does not prove that the filaments are actually coherent structures that trace the ``X'' shape seen clearly at lower latitudes. The average surface brightness of the NE box (without subtracting background) is 20.4~mag~arcmin$^{-2}$, which is about equal to the 1$\sigma$ contour above background. This precludes definitively characterizing the emission as a coherent filament. However, there are two reasons to believe that they do continue the ``X'' shape. 

First, the X-ray image has similar structures that are clearly connected to the lower latitude filaments. This also makes it unlikely that they are filter artifacts. There are some \Galex\ FUV images where either a ghost image or stray light from a bright source exterior to the field causes similar looking patterns, but there are no large-scale structures like this seen in the wider FUV or NUV images. 

Second, an azimuthal map gridded in 15$^{\circ}$ angular segments at several radii (based on the drop-off of the average radial profile) suggests that the ``X''-shaped structure continues beyond the inner regions except to the southwest. The right panel in Figure~\ref{fig:fuv_filaments} shows the quantity $(X-\mu)/\sigma$, where $X$ is the mean value of the FUV image in each segment, $\mu$ is the mean of the $X$ values in that ring, and $\sigma$ is their standard deviation (\textit{not} the background). There are no filament candidates exceeding 3$\sigma$ from the mean at any radii, but note that if the ``X'' shape is real then at least four of the 24 segments will have positive deviations. For instance, if the background annulus has a uniform brightness and there are four equally brighter segments, the limiting $(X-\mu)/\sigma \approx 2.2$. Therefore, the purpose is to show where positive deviations occur, and the map shows that they form an ``X'' shaped structure in the second ring ($\theta=1.25$ to 5~arcmin) and this continues to the northwest, northeast, and southeast in the outer ring ($\theta=5$ to 11.25~arcmin). 

% RA Dec Angle Length Width Flux Luminosity 
\begin{deluxetable*}{cccccccccc}
\tablenum{2}
\tabletypesize{\scriptsize}
\tablecaption{\label{table.filaments} UV Filament Luminosities}
\tablewidth{0pt}
\tablehead{
\colhead{(1)} & \colhead{(2)} & \colhead{(3)} & \colhead{(4)} & \colhead{(5)} & \colhead{(6)} & \colhead{(7)} & \colhead{(8)} & \colhead{(9)} & \colhead{(10)} \\
\colhead{Filament} & \colhead{R.A.} & \colhead{Dec} & \colhead{P.A.} & \colhead{Length} & \colhead{Width} & \colhead{Filter} & \colhead{$m$} & \colhead{FUV$-$NUV} & \colhead{$\lambda L_{\lambda}$} \\
  & \colhead{(J2000)} & \colhead{(J2000)} & \colhead{(deg.)} & \colhead{(arcmin)} & \colhead{(arcmin)} &  & \colhead{(mag)} & \colhead{(mag)} & \colhead{($10^{40}$~erg~s$^{-1}$)}
}
\startdata
NE & 10:02:32.16 & $+$55:47:15.8 & 38  & 7 & 2.4 & FUV & 19.8$\pm$0.1   & $< -0.4$     & 3.7$\pm$0.4 \\
   &             &               &     &   &     & NUV & $>$20.2        &              & $<1.8$ \\ 
SE & 10:02:38.20 & $+$55:37:37.1 & 110 & 7 & 2.4 & FUV & 19.6$\pm$0.1   & $< -0.7$     & 4.6$\pm$0.4 \\
   &             &               &     &   &     & NUV & $>$20.3        &              & $<$1.6 \\ 
NW & 10:01:15.92 & $+$55:43:11.2 & 110 & 7 & 2.4 & FUV & 19.8$\pm$0.1   & $<-0.6$      & 3.7$\pm$0.3 \\
   &             &               &     &   &     & NUV & $>$20.4        &              & $<$1.5\\ 
SW & 10:01:24.17 & $+$55:35:57.9 & 60  & 7 & 2.4 & FUV & 20.4$\pm$0.2   & $< 0.1$      & 2.1$\pm$0.4 \\
   &             &               &     &   &     & NUV & $>$20.3        &              & $<$1.3 \\ 
\enddata
\tablecomments{Cols. (1) Filament ID from Figure~\ref{fig:fuv_filaments} (2-3) Central coordinates (4-6) Region (7) GALEX filter (8) AB magnitude calculated from the GALEX count rate after PSF wing and dust correction (9) FUV-NUV color (10) Luminosity}
\end{deluxetable*}

\begin{figure}
    \centering
    \includegraphics[width=0.5\textwidth]{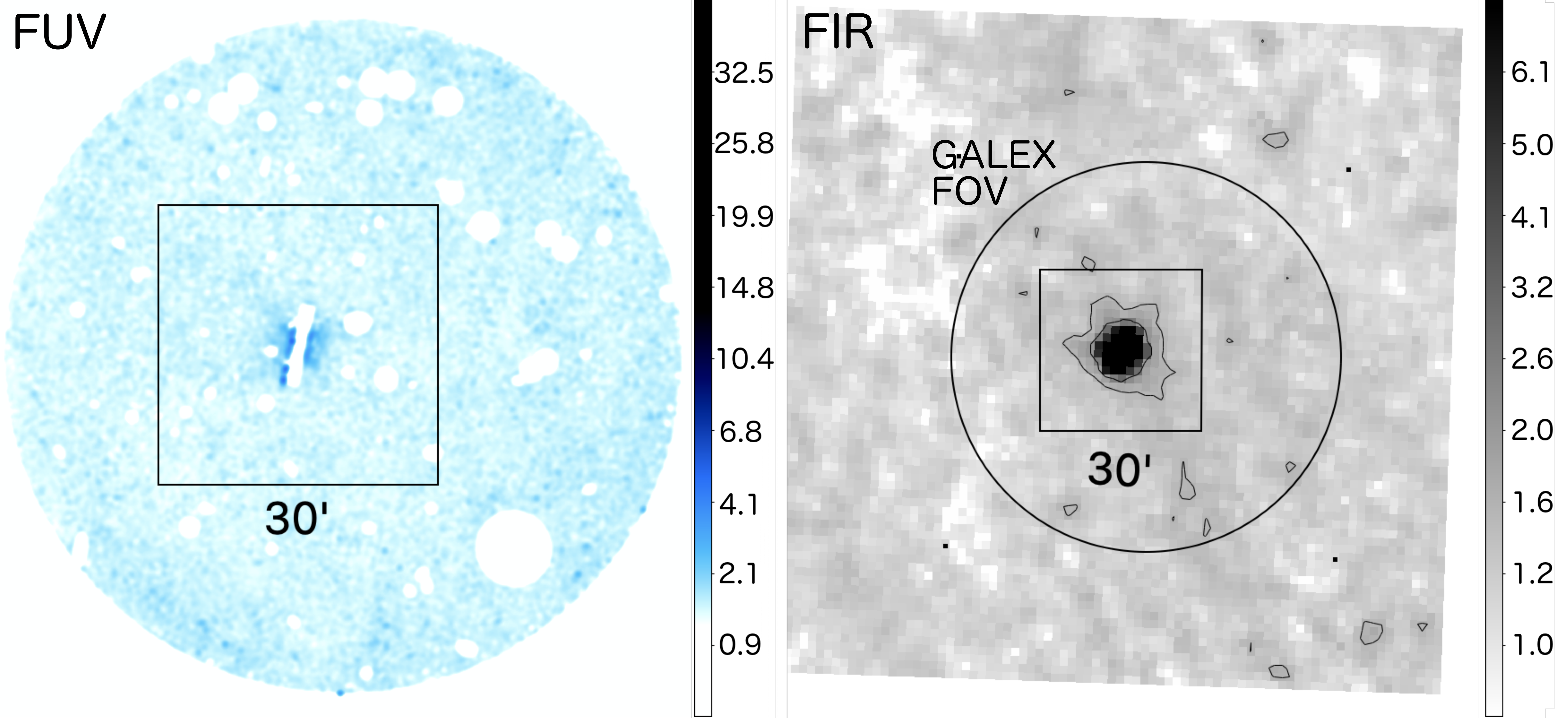}
    \caption{Neither the full \Galex\ FUV image (\textit{left}) nor the 100~$\mu$m IRAS image (\textit{right}) support the hypothesis that the filaments are Galactic cirrus. The cirrus in the region of NGC~3079 is neither bright nor strongly spatially variable. Since the galaxy itself is a bright point source at 100~$\mu$m (89~Jy), it is not possible to search for FIR counterparts in the IRAS data. The box shows the region shown in Figure~\ref{fig:images}, while the circle shows the \Galex\ field of view. The FUV color scale is in units of $10^{-4}$~counts~s$^{-1}$, while the IRAS color scale is in units of MJy~sr$^{-1}$. Sources detected with greater than 3$\sigma$ significance have been masked in the FUV image.}
    \label{fig:cirrus}
\end{figure}

\subsection{Galactic cirrus}

Extensive Galactic cirrus filaments are common in \Galex\ images, so we investigated whether the filaments around NGC~3079 can be explained by cirrus. First, we searched for larger structures in the full 2$^{\circ}$ FUV image (as well as in the NUV image). There is no clear structure in this wider field in either the FUV (Figure~\ref{fig:cirrus}) or the NUV. The FUV$-$NUV colors of less than $-0.4$~mag in the filament boxes shown in Figure~\ref{fig:fuv_filaments} also disfavor cirrus, which tends to have FUV$-$NUV$ \sim 0$. In this case, we should have easily detected the filaments in the NUV image. 

We then searched for far-infrared (FIR) counterparts, or evidence for filamentary or highly variable FIR emission from cirrus in the region. To investigate the larger-scale emission, we used the IRAS 100~$\mu$m maps\footnote{retrieved from https://irsa.ipac.caltech.edu/data/IRIS/}, as reprocessed using the IRIS software \citep{Miville-Deschenes2005}. Figure~\ref{fig:cirrus} shows the wider field in the FUV and in the FIR, and the cirrus in the region appears somewhat patchy but with variation on larger scales than the filaments. NGC~3079 itself is a FIR source, and the 4\arcmin\ resolution of the IRAS map prevents directly searching for FIR counterparts to the FUV filaments. However, no such counterparts are seen in the higher resolution, but smaller field of view, \Herschel\ images of the region (covering 70-200~$\mu$m). Since diffuse FIR emission from dust above NGC~3079 itself (``cirrus'' in that galaxy) is detected at lower latitudes in the same maps, we estimate that FIR emission from cirrus would be detectable considering the FUV fluxes. Finally, we examined the fields around NGC~3079 that have shallow \Galex\ FUV coverage. We do not find such sharply defined structure within a few degrees of NGC~3079. Thus, we argue that it is very unlikely that the large FUV filaments are due to Galactic cirrus.

They are also not due to scattered light from dust above NGC~3079 itself. At lower latitudes, dust scattering is evident through UV continuum emission in the FUV, NUV, and \Swift\ UVOT bands \citepalias{Hodges-Kluck2016}. The FUV$-$NUV color of the diffuse, low latitude scattered light is close to 0~mag, whereas FUV$-$NUV$ < -0.4$~mag for three of the four filaments. 

If the FUV filaments are not produced by dust in either the Galaxy or NGC~3079, the most likely light source is emission lines in the filter. Any other continuum source would have to explain the absence of emission in redder bands. There are several possible lines covered by the FUV filter, including \ion{C}{4} $\lambda\lambda$1548, 1550\AA, \ion{He}{2} $\lambda$1640\AA, and [\ion{O}{3}] $\lambda\lambda$1661, 1666\AA. These lines could be produced either by cooling wind fluid, photoionization, or shock heating. 

Without spectroscopic data, it is not possible to use the FUV fluxes alone to distinguish between these possibilities. If such a spectrum were to exist, models such as {\sc cloudy} \citep{Ferland2017} would be able to do so. For example, the \ion{He}{2} line requires hard ionizing photons (at least 54~eV), so a strong \ion{He}{2} line would favor shocks or AGN photoionization \citep[cf.][]{Jaskot2016}. As far as we know, there are no existing models that could use only the FUV flux to determine the origin, but we note that \citet{Borthakur2013} found \ion{C}{4} \textit{absorption} around four starburst galaxies at impact parameters between 100-200~kpc and concluded from {\sc cloudy} modeling that photoionization from either a starburst or the metagalactic radiation field cannot explain the \ion{C}{4} lines. They instead find that shock heating can do so, provided that much of the wind energy is expended in shock-heating circumgalactic gas. The emission around NGC~3079 is not as extended, but if it connects to unseen ionized gas at larger radii then the same logic would apply.

\subsection{Quadrant Stacks}

\begin{figure*}
    \centering
    \hspace{-0.55cm}\includegraphics[height=2.2in,trim=60 340 -10 160,clip]{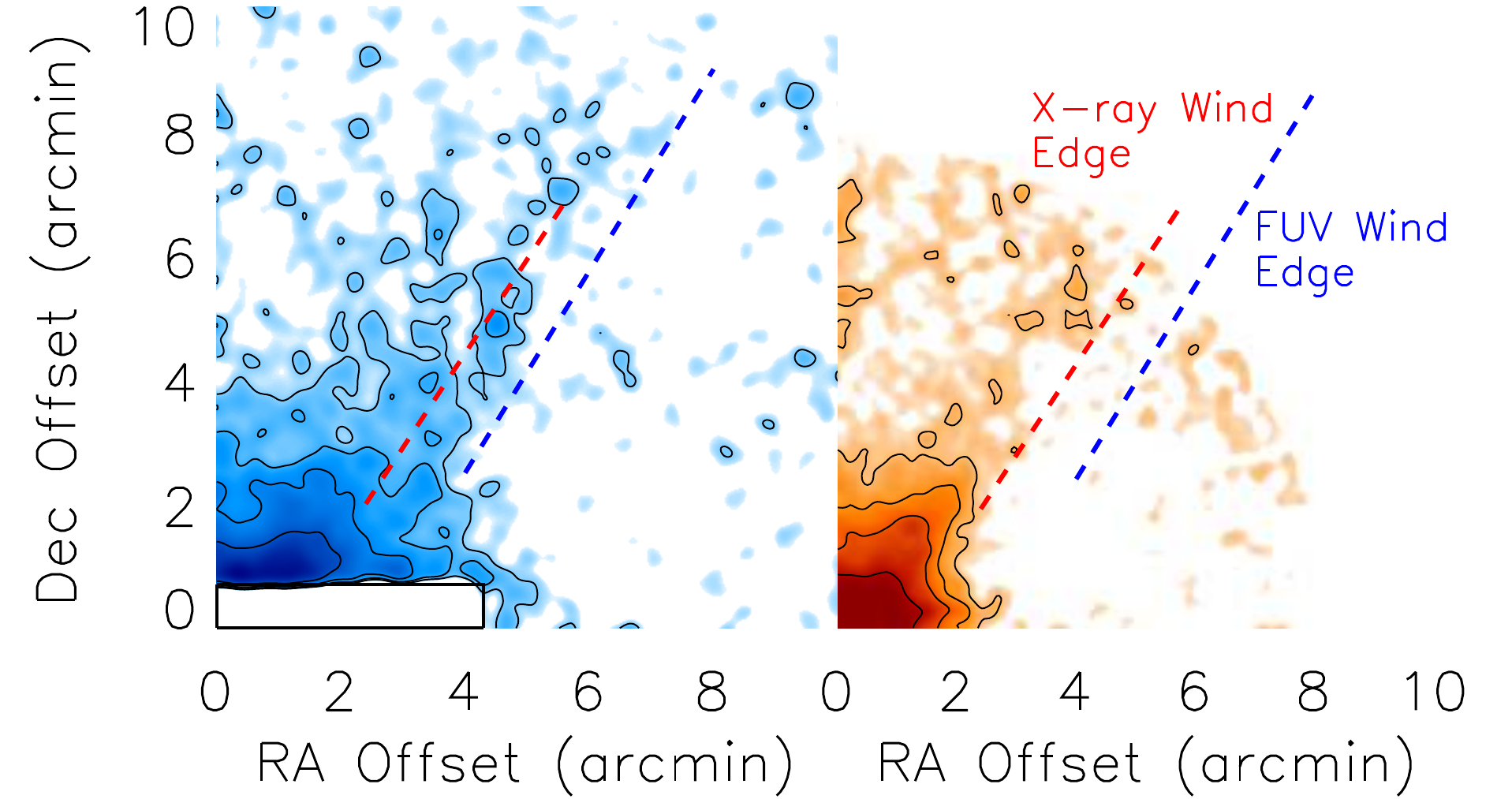}
    \includegraphics[height=2.2in]{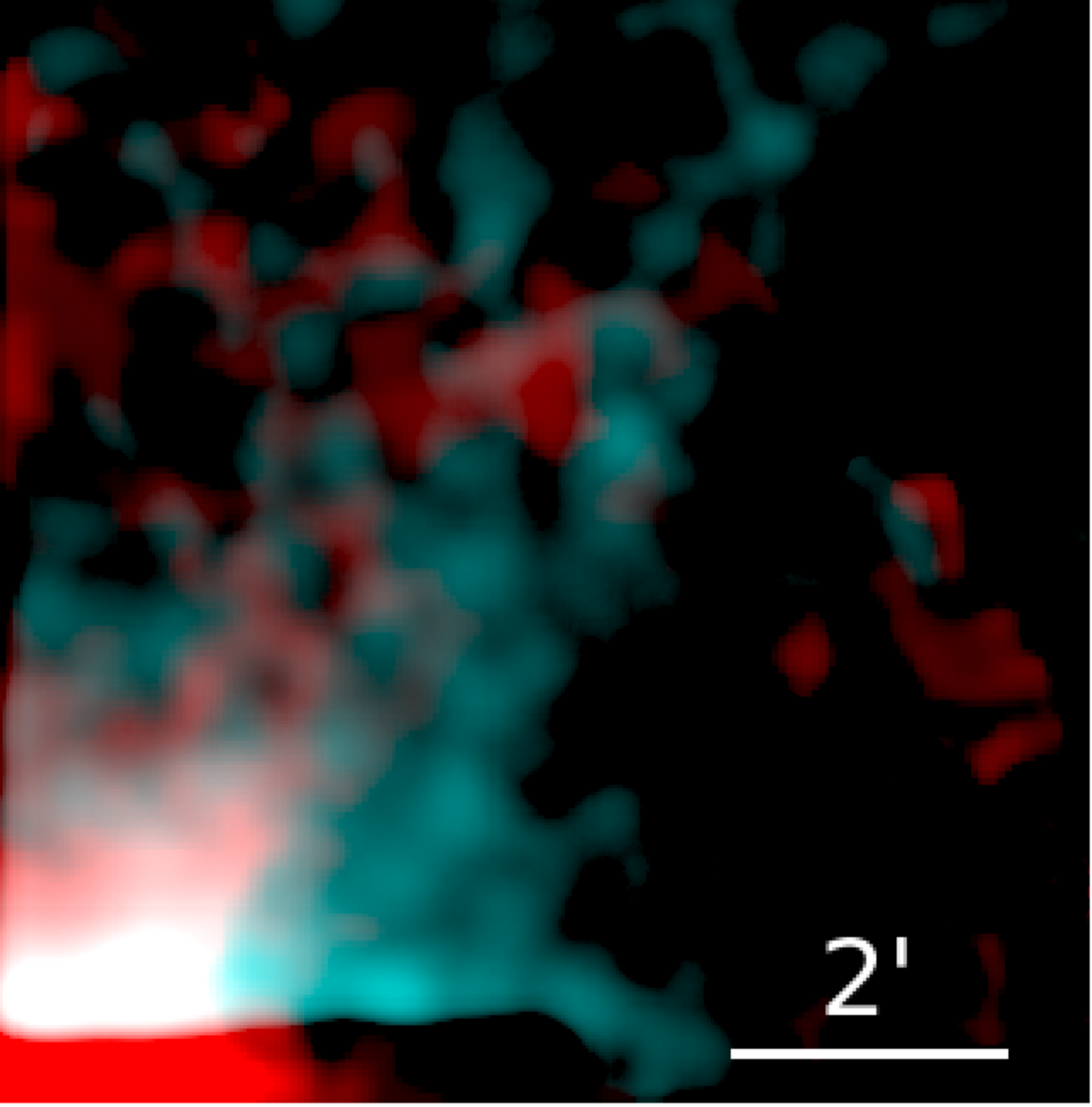}
    \caption{FUV (\textit{left}) and X-ray (\textit{center}) images rotated and stacked into a single quadrant to improve the signal, with the galaxy nucleus at the origin and the galactocentric radius along the $x$-axis. The images have been clipped at the mean background, smoothed, and overlaid with 2, 4, 8, and 12$\sigma$ contours. The X-ray image is clipped at a radius of 8$^{\prime}$ because of the lower signal at the image edges. Both images show an extended, conical, limb-brightened wind profile, with the FUV wind extending farther than the X-ray wind in both the horizontal and vertical directions. 
    \textit{Right}: The X-ray (red) and FUV (cyan) quadrant images are clipped at the mean background + 1$\sigma$ and overlaid. The combination highlights that the X-rays are interior to the FUV emission.}
    \label{fig:quad_plot}
\end{figure*}

\begin{figure}
    \centering
    \includegraphics[trim=60 380 10 130,clip,width=0.5\textwidth]{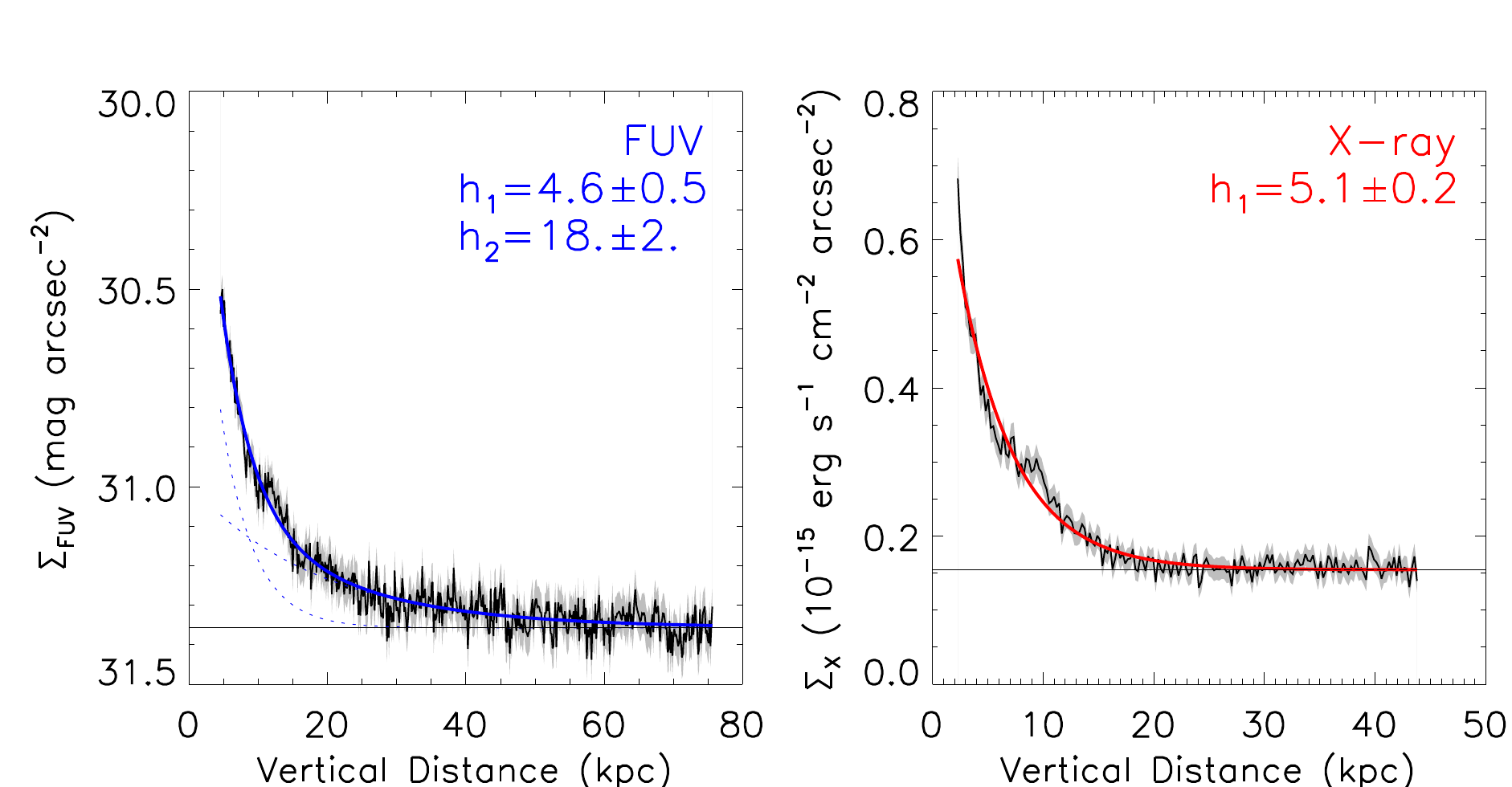}
    \includegraphics[trim=60 380 10 130,clip,width=0.5\textwidth]{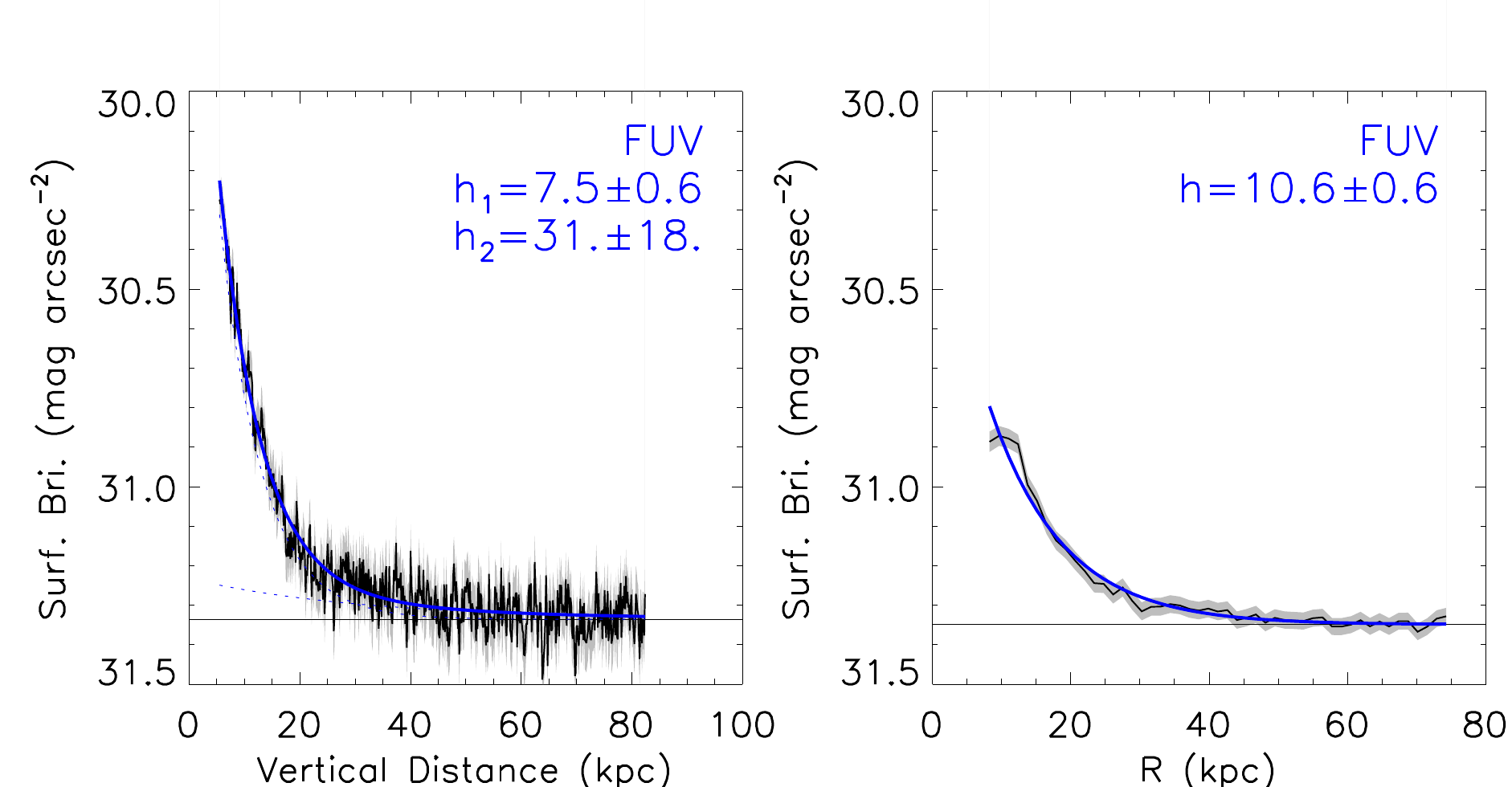}    
    \caption{Emission profiles measured from the quadrant stacks (Figure~\ref{fig:quad_plot}), with the best-fit scale heights listed. Scale heights for both components are listed when a double-exponential model is a significantly better fit than a single-exponential model. The data are shown in black (errors in grey) and the best-fit model plotted in blue (FUV) or red (X-rays), with the components plotted as dotted lines. Here we show the FUV vertical ($z$) profile averaged over $R<22$~kpc (\textit{top left}), the X-ray vertical profile averaged over $R<16$~kpc (\textit{top right}), the FUV profile along the wind cone edge as shown in Figure~\ref{fig:quad_plot} (\textit{bottom left}), and the FUV radial profile (\textit{bottom right}). The large scale heights indicate that the wind is indeed very extended. }
    \label{fig:quad_prof}
\end{figure}

The structure of the FUV and X-ray emission becomes clearer when stacking them in quadrants (Figure~\ref{fig:quad_plot}). We define the abscissa along the galactocentric radius $R$ and the height above the midplane $z$ as the ordinate. Despite the uncertainty about the filaments, the FUV and X-ray emission clearly trace limb-brightened cones. Notably, the edge of the X-ray cone is interior to that of the FUV cone, and the X-ray edge coincides with the brightest part of the FUV limb. This morphology is consistent with the hot wind model from \citet{Heckman1990}, in which the hot wind shocks surrounding gas, and this shocked gas is surrounded by a thin layer of rapidly cooling gas visible in optical and UV emission lines. On the other hand, the morphology disfavors strong radiative cooling of the hot wind fluid itself \citep{Thompson2016}, in which case we would expect strongly peaked X-ray emission below the strongest FUV line emission. 

We measured scale heights for several types of profiles. These include vertical profiles, radial profiles, and profiles extracted along the edge of the cone (Figure~\ref{fig:quad_prof}). The vertical profiles are averaged within $R<22$~kpc for the FUV image and $R<16$~kpc for the X-ray image, based on the extent of the base of the wind. A single exponential profile fits the X-ray data well but not the FUV, whereas the sum of two exponential profiles is a good fit to the FUV data. The scale height in the X-rays is $h_X = 5.1\pm0.2$~kpc, which is consistent with one component of the FUV, $h_{\text{FUV},1} = 4.6\pm0.5$~kpc. However, the second component of the FUV, which accounts for 50\% of the flux, has a much larger scale height of $h_{\text{FUV},2} = 18\pm2$~kpc. Meanwhile, the FUV profile along the edge of the wind cone (where the $S/N$ is highest) is even more extended, with scale heights $h_{\text{FUV,1}} = 7.5\pm0.6$~kpc and $h_{\text{FUV},2} = 31\pm18$~kpc, while the azimuthally averaged radial profile has $h_{\text{FUV}} = 10.6\pm0.6$~kpc. 

% Height (arcmin) Height (kpc) A (arcmin) kT (keV) norm (1e-6) LX | R2 R1 n*Z/Zodot mass
\begin{deluxetable*}{cccccc|ccccc}
\tablenum{3}
\tabletypesize{\scriptsize}
\tablecaption{X-ray Wind Properties}
\tablewidth{0pt}
\tablehead{
\colhead{(1)} & \colhead{(2)} & \colhead{(3)} & \colhead{(4)} & \colhead{(5)} & \colhead{(6)} & \colhead{(7)} & \colhead{(8)} & \colhead{(9)} & \colhead{(10)} & \colhead{(11)} \\
\colhead{Height} & \colhead{Height} & \colhead{Area} & \colhead{$kT$} & \colhead{{\sc apec} norm} & \colhead{0.3-10 keV $L_{X}$} & \colhead{$R_2$} & \colhead{$R_1$} & \colhead{$h$} & \colhead{$\bar{n}(Z/Z_{\odot})$} & \colhead{Hot Mass}\\
\colhead{(arcmin)} & \colhead{(kpc)} & \colhead{(arcmin$^2$)} & \colhead{(keV)} & \colhead{($10^{-5}$)} & \colhead{($10^{39}$ erg~s$^{-1}$)} & \colhead{(arcmin)} & \colhead{(arcmin)} & \colhead{(arcmin)} &  \colhead{($10^{-3}$ cm$^{-3}$)} & \colhead{($10^6 M_{\odot}$)}
}
\startdata
-3.9    & -21.5 & 12.5  & 0.28$\pm$0.12 & $<$0.2                 & $<$0.2              & 3.1$\pm$0.3 & 2.6$\pm$0.8 & 1.6 & $<$0.15 & $<$5\\
-2.4    & -13.2 & 9.3   & 0.28$\pm$0.02 & 0.80$^{+0.06}_{-0.09}$ & 0.7$\pm$0.1         & 1.9$\pm$0.2 & 1.5$\pm$0.4 & 1.2 & 0.6$\pm$0.2 & 8$\pm$3 \\
-1.2    & -6.6  & 9.5   & 0.36$\pm$0.02 & 2.4$\pm$0.1            & 2.4$\pm$0.1         & 1.4$\pm$0.2 & 1.1$\pm$0.4 & 1.2 & 1.4$\pm$0.3 & 10$\pm$3 \\
0       & 0     & 8.0   & 0.40$\pm$0.01 & 6.8$\pm$0.4            & 4.8$\pm$0.3         & 1.1$\pm$0.3   & 0.5$\pm$0.2   & 1.2 & 2.0$\pm$0.6 & 18$\pm$7 \\
1.2     & 6.6   & 9.5   & 0.33$\pm$0.03 & 2.2$^{+0.2}_{-0.1}$    & 2.1$^{+0.2}_{-0.1}$ & 1.4$\pm$0.2 & 1.2$\pm$0.4 & 1.2 & 1.3$\pm$0.3 & 9$\pm$2 \\
2.4     & 13.2  & 9.7   & 0.29$\pm$0.02 & 0.9$\pm$0.1            & 0.8$\pm$0.1         & 2.0$\pm$0.2 & 1.5$\pm$0.3 & 1.2 & 0.6$\pm$0.2 & 7$\pm$2 \\
3.9     & 21.5  & 13.8  & 0.27$\pm$0.02 & 0.5$\pm$0.1            & 0.4$\pm$0.1         & 3.5$\pm$0.3 & 2.1$\pm$0.6 & 1.6 & 0.3$\pm$0.1 & 11$\pm$5
\enddata
\tablecomments{\label{table:xrayprop} Cols. (1-2) Average box height above midplane (3) Unmasked area for spectral extraction (4-6) Best parameters for an isothermal model (7-9) Best parameters for the average cylindrical shell at each height (10) Mean density in each shell. The errors assume a 30\% error in $R_1$, except for the central aperture where we assume a 50\% error because source masking makes it more difficult to determine ($R_2$ is well defined and $h$ is fixed). The error bar for $kT$ in the case of the density upper limit corresponds to allowed temperatures if the source is there. 
}
\end{deluxetable*}

\section{X-ray Properties}
\label{section.xrays} 

\begin{figure}
    \centering
    \vspace{0.25cm}
    \includegraphics[width=0.47\textwidth]{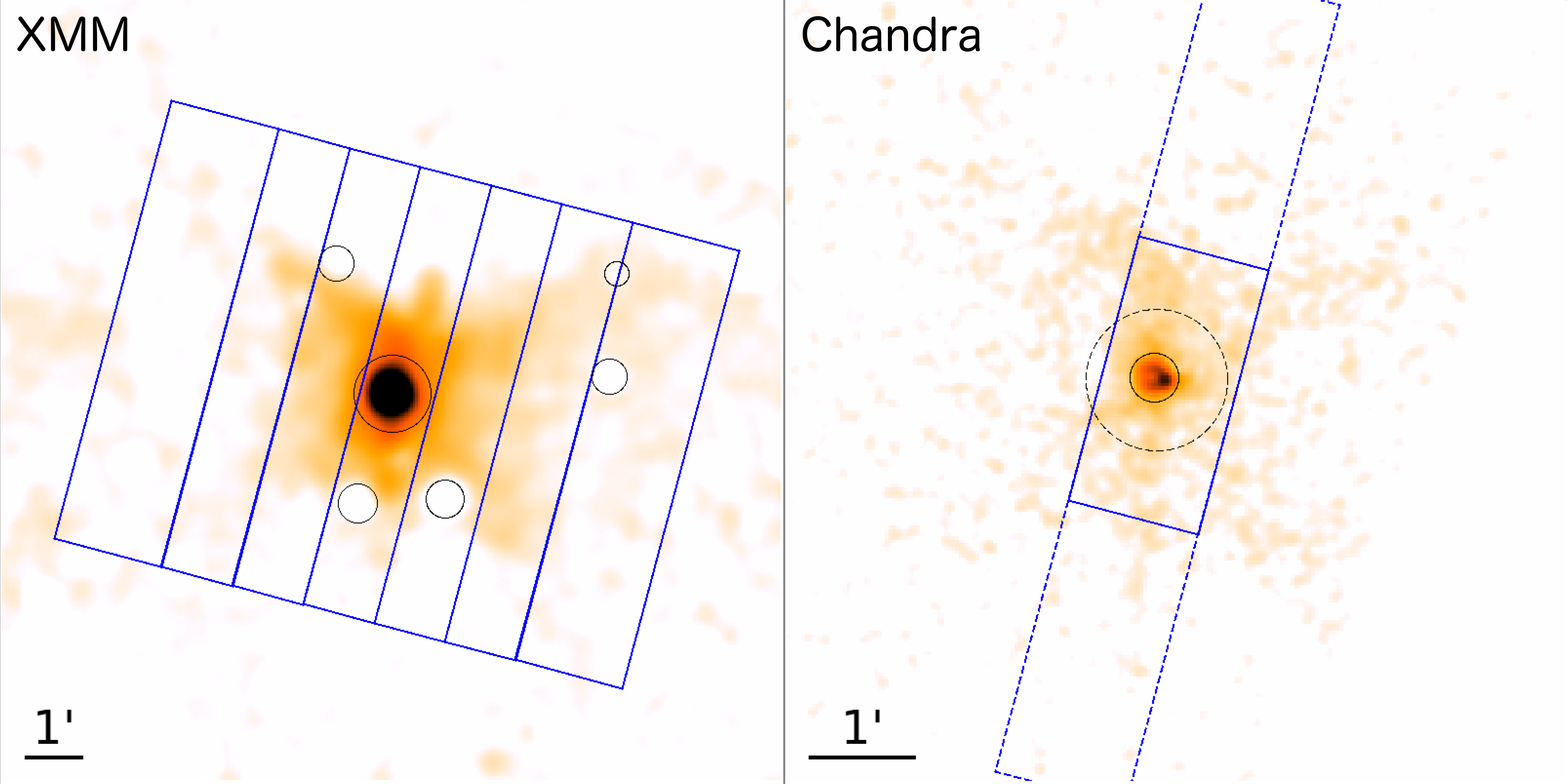}
    \includegraphics[trim=50 380 10 100,clip,width=0.47\textwidth]{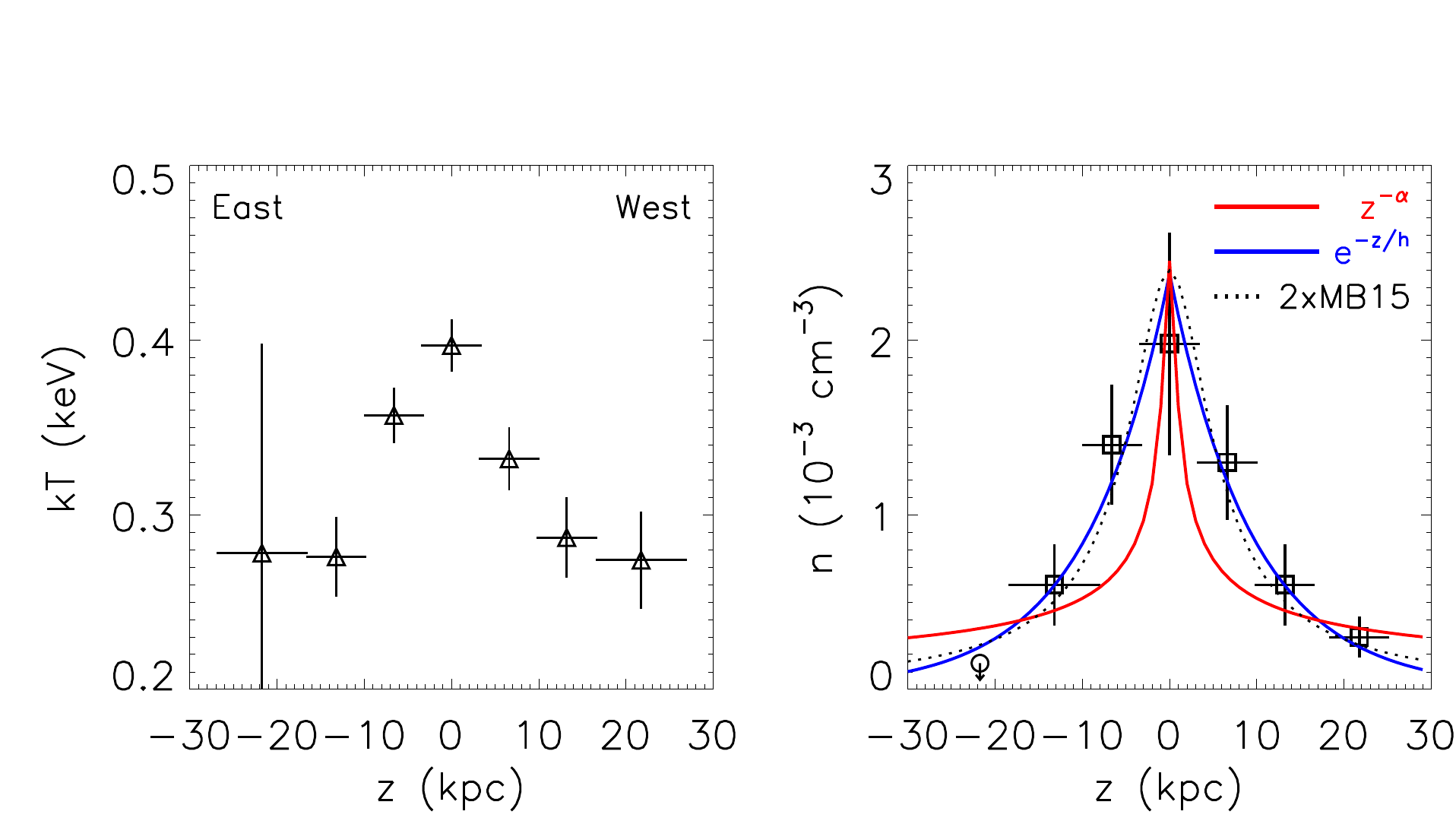}
    \caption{\textit{Top Left}: The temperature and density were inferred from model fits to spectra extracted from the blue rectangular regions, shown overlaid on the soft X-ray image with point source masks marked in black. \textit{Top Right}: The bright superbubble is unresolved with XMM and its mask (dashed circle) covers much of the central region, so we measured the flux and emission measure from the \Chandra\ image where the bubble (solid circle) is resolved (see text). \textit{Bottom Left}: The temperature profile measured from these spectra, based on fitting isothermal plasma models to the data. The error bars are the 90\% credible interval. \textit{Bottom Right}: Mean density inferred from the emission measure and a cylindrical wind model (see text). The red and blue lines are the best-fitting power-law and exponential disk models, respectively, while the dotted line represents a shock compression of the Milky Way halo model \citep{Miller2015} by a factor of two.
    }
    \label{fig:tprof}
\end{figure}

The soft X-rays provide further insight into the wind through the resolved temperature and density measurements. We measured these from spectra extracted in apertures parallel to the midplane (Figure~\ref{fig:tprof}). The vertical sizes of the apertures are based on the need to obtain at least several hundred source counts to measure precise temperatures through most of the wind. There is more than enough signal at lower heights, where both soft and hard X-rays have been previously studied \citep{Strickland2004,JiangtaoLi2019}, to perform finer gridding, but we defer a more complete analysis of the temperature structure to a future paper (Yukita et al., in prep.). 

\subsection{Spectral Extraction and Fitting}

For each aperture (Figure~\ref{fig:tprof}), we masked point sources and extracted spectra from each XMM observation using the XMM-ESAS software \citep{Snowden2004}. The bright superbubble at the center \citep{Cecil2002} is point-like at the XMM resolution and is also masked. An inspection of the Chandra image shows that the diffuse spectrum from the central aperture predominantly contains flux from the base of the wind, so we include that aperture in the profiles. We also extracted a background spectrum from an annular aperture for each exposure. 

All spectral fitting was performed with \textit{Xspec} v12.10.1, and the spectra from a given region were fitted jointly rather than using a co-added spectrum. We used the {\sc apec} thermal plasma model with the \citet{Asplund2009} solar abundance table to model hot gas. If the X-rays come from shocked gas, {\sc apec} may not be appropriate as it is based on collisional ionization equilibrium. However, the equilibrium timescales are short relative to both the radiative lifetime and the time for the wind to expand at least to 50~kpc, so the hot gas at the wind edge is likely close to equilibrium.   

Instead of subtracting a scaled background spectrum, we fitted the background spectrum and used the best-fit model as a fixed component when fitting the source spectra. We adopted a model for the soft X-ray background that includes a thermal component for the Local Hot Bubble and the Galactic hot halo, and a power law with photon index $\Gamma = 1.46$ for the cosmic X-ray background ({\sc phabs(apec+apec+pow)}), in addition to instrumental lines and continua\footnote{described in the ESAS manual and particular to each observation; see https://heasarc.gsfc.nasa.gov/docs/xmm/esas/cookbook/xmm-esas.html}. We obtained a good fit with typical values ($kT_1 = 0.1$~keV and $kT_2 \approx 0.25$~keV). 

The source model is an absorbed, isothermal plasma ({\sc phabs*apec}), where the column density of the absorbing material is fixed at the Galactic value $N_{\text{H}} = 9\times 10^{19}$~cm$^{-2}$. We fix the metallicity ($Z/Z_{\odot}$) at the solar value for these fits, but the signal is sufficient to measure the O/Fe ratio (although not the abundances of individual elements), as explored below.

\begin{figure}
    \centering
    \includegraphics[trim=70 380 50 70,clip,width=0.47\textwidth]{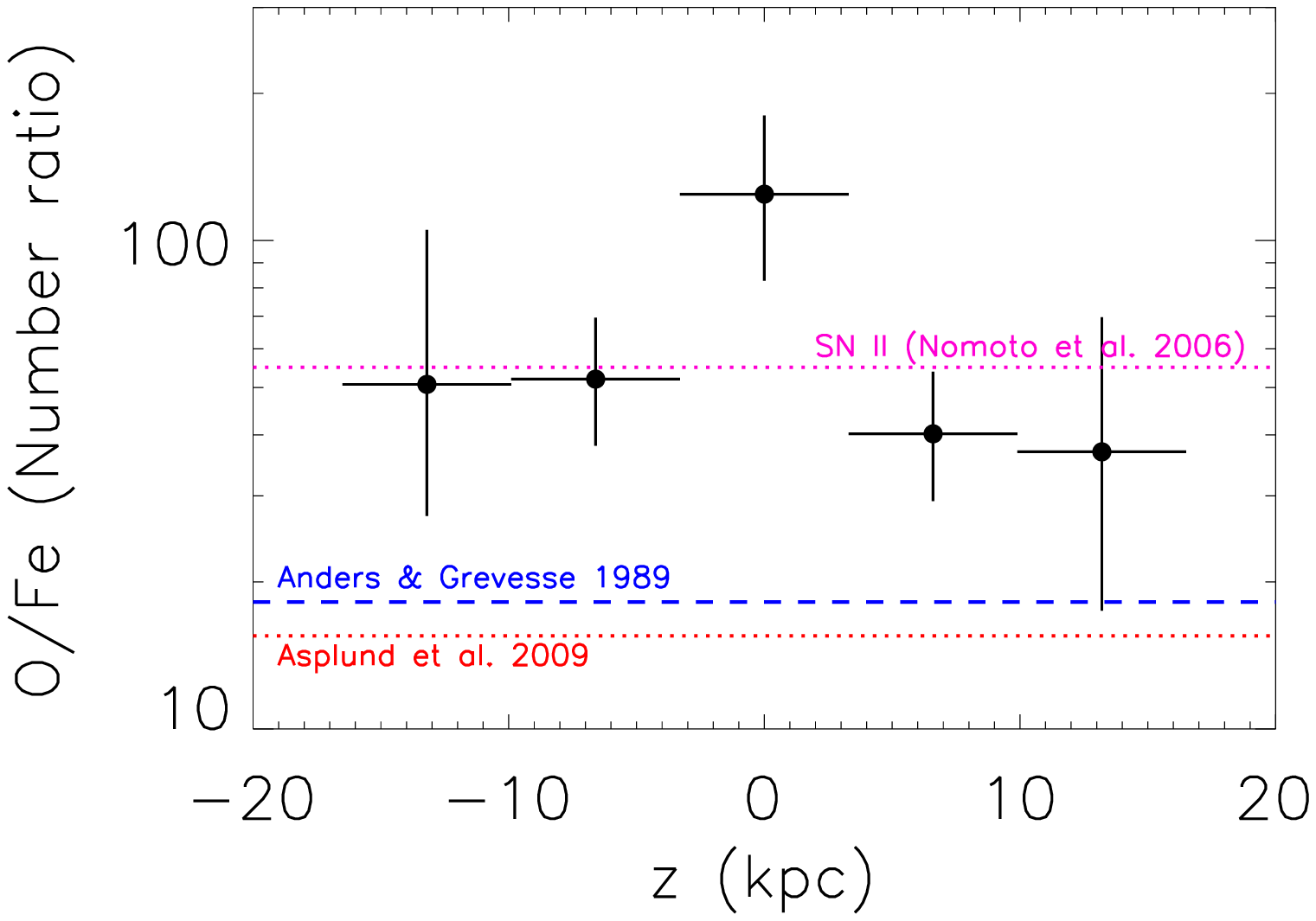}
    \caption{
    O/Fe number density as a function of height above the midplane. The O and Fe values are measured using an isothermal {\sc vapec} model using \citet{Asplund2009} for solar abundances. The outermost bins are not shown because the signal in the spectrum is too low to measure abundances. The expected ratios for Type~II SNe from \citet{Nomoto2006} and 
    the Sun \citep{Anders1989, Asplund2009} are shown for reference. The outflowing material is more consistent with enrichment from Type~II SNe, in agreement with a prior Suzaku study by \citet{Konami2012}.
    }
    \label{fig:abundprof}
\end{figure}

\subsection{Temperature, Density, and Abundance Profiles}

The best-fit temperatures, {\sc apec} normalizations, and luminosities are given in Table~\ref{table:xrayprop}. The X-ray surface brightness is asymmetric from east to west across the disk, possibly because of the interaction of the wind with companion galaxies on the west side. There is insufficient signal in the easternmost bin to robustly measure the temperature, although the wind is formally detected. 

The {\sc apec} normalization encodes the emission measure, $\text{EM} \propto \int n_e n_{\text{H}} dV$. To obtain $\bar{n}$ we assume a volume and that $n_e = n_{\text{H}}$. Based on the limb-brightened morphology, in each aperture except at the midplane we fitted a simple cylindrical shell model with constant density in the shell to the surface brightness to obtain the emitting volume. This model yields the average inner and outer radii $R_1$ and $R_2$, while the height $h$ is that of the aperture. Since we have fixed the metallicity at Solar, we derive the quantity $\bar{n} (Z/Z_{\odot})^{1/2}$. A summary of the measured properties is given in Table~\ref{table:xrayprop} and the temperature and density profiles are shown in Figure~\ref{fig:tprof}. 

The central aperture is a special case because the XMM superbubble mask covers much of the emission of interest (Figure~\ref{fig:tprof}). Thus, to estimate the density we first obtain $R_1$, $R_2$, and the temperature from the XMM data. Then, we measure the $0.3-2$~keV Chandra count rate in the region using a more appropriately sized mask for the superbubble, whose emission is associated with radio lobes and is not part of the larger X-shaped structure \citep{Irwin2003}. The Chandra response files are used to convert this count rate into an emission measure, assuming the best-fit XMM {\sc apec} temperature, and the emission measure is converted to density using the same assumptions as for the other apertures.  

The profiles indicate a decline in temperature as well as density. The temperature appears to flatten near $kT \sim 0.28$~keV, but in the outer regions the signal is low and the fit may be biased by the Galactic background, which has a similar $kT \sim 0.25$~keV. The decline of the density is more certain. The inferred density is weakly sensitive to the volume assumed in the shell model, but $R_2$ is tightly constrained to within 1.5~kpc. $R_1$ is less constrained, but a filled cylindrical model is strongly ruled out by the limb brightening and the uncertainty in the quantity $R_2^2-R_1^2$ is dominated by that in $R_2$. 

The density profile cannot be described by a power-law model, where $n(z) \propto z^{-\alpha}$ (Figure~\ref{fig:tprof}). The best-fit $\alpha = 0.26\pm0.05$ is obviously a poor fit to the data. An exponential profile, where $n(z) \propto e^{-z/h}$, is a good match with a scale height $h=10\pm3$~kpc.  This is consistent with the $\approx$5~kpc scale height from the surface brightness profile, which scales as $n^2$. For a symmetric biconical wind with an opening angle of 30$^{\circ}$, the best-fit profiles imply a total mass within 30~kpc of $\approx 7\times 10^7 M_{\odot}$ for the X-ray-emitting gas.

Finally, Figure~\ref{fig:abundprof} shows the O/Fe ratio as a function of height above the midplane. This ratio was determined by fitting the spectra in each box with an isothermal {\sc vapec} model in which the abundances of the $\alpha$ elements and Fe-group elements were allowed to float relative to the solar values. The number ratio was then calculated using the solar abundance table, for which we adopted that of \citet{Asplund2009}. We then compared the values to the predictions for Type~II SNe enrichment \citep{Nomoto2006} and the solar abundance. The measured values are a better match to the Type~II SNe, indicating that the outflow is enriched and powered by these SNe.

There are several caveats. First, for the temperature of NGC~3079 and the energy resolution of the MOS and pn, the oxygen abundance is the best cosntrained and the difference between \citet{Asplund2009} and other commonly used tables is significant for oxygen. However, using another table with {\sc vapec} \citep[such as][]{Anders1989} would tend to increase the O/Fe ratio rather than bring it closer to the Solar value. Second, the signal in the spectra is too low to rule out solar abundances for regions well above the disk. Indeed, the outermost bins lack the signal to get a meaningful fix on O/Fe at all and are not shown. Third, if the wind has a complex temperature structure then the abundance values are likely biased. We will explore these issues in a forthcoming, detailed analysis of the X-ray spectra (Yukita et al., in prep.).

\begin{figure*}
    \centering
    \includegraphics[height=5.7cm]{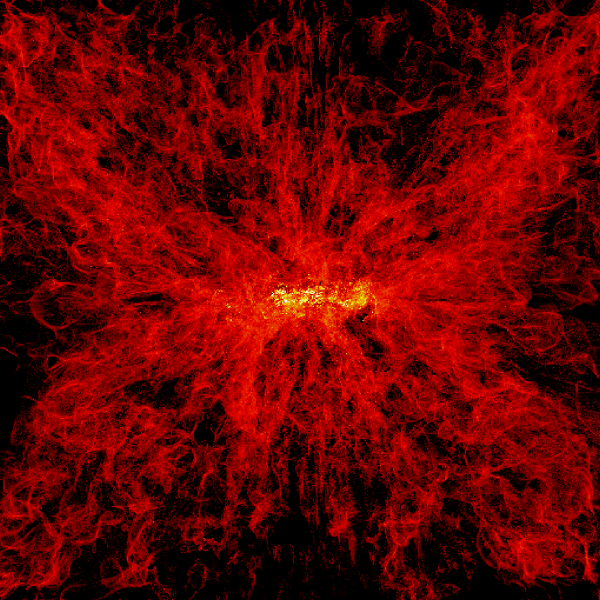}
    \includegraphics[height=5.7cm]{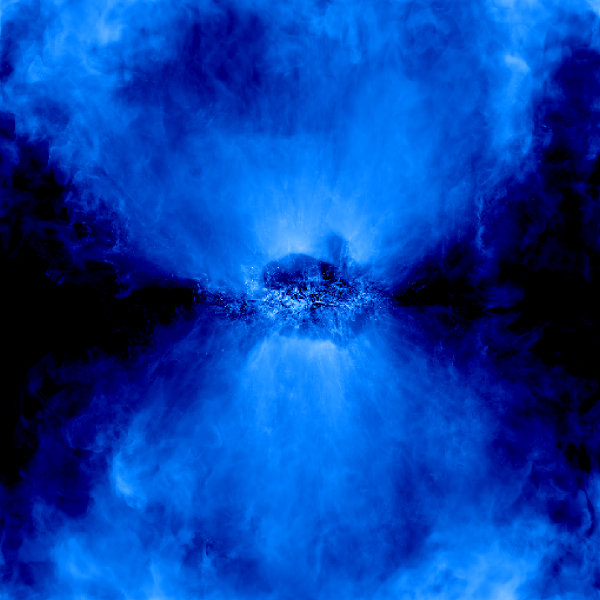}
    \includegraphics[height=5.7cm]{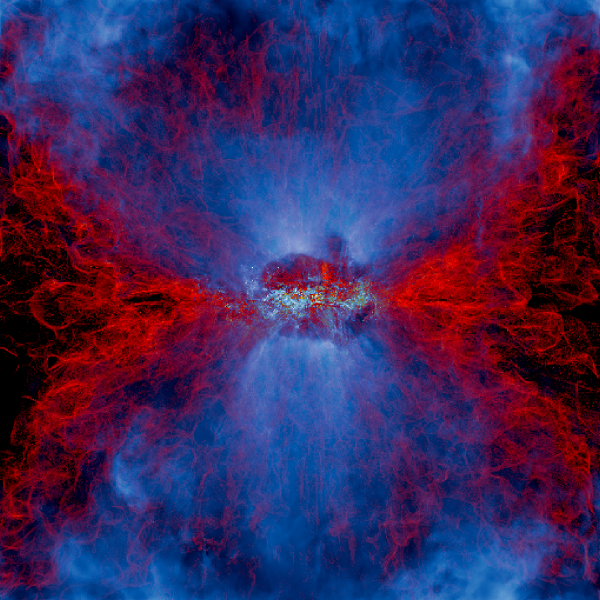}
    \caption{Projected emission from a simulated galactic wind from \citet{Tanner2016}. The three panels show H$\alpha$ emission (left), which has the same morphology as FUV line emission, soft X-ray emission (center) from the shocked wind, and the superposition of X-rays on H$\alpha$ (right). Note that the X-rays are somewhat edge-brightened, mostly ``fill'' the wind cone at lower latitudes, and are interior to the strongly edge-brightened H$\alpha$ emission, similar to what is seen in NGC~3079. ach box is 1~kpc on a side, but see text for discussion.}
    \label{fig:simulations}
    \vspace{0.25cm}
\end{figure*}

\section{The Extended Wind}
\label{section.interpretation}

NGC~3079 is a well known superwind galaxy and the features described above are new only in their extent, as they continue the X-shaped structure previously reported \citep{Heckman1990,Cecil2001}. Thus, we conclude that the X-ray, and especially FUV, emission are part of an extended galactic wind on a scale similar to that seen in the galaxy Makani by \citet{Rupke2019}. This extended emission provides clues to the formation of the wind, for which there are multiple theories \citep[reviewed by][]{Zhang2018}. 

In particular, the line emission (primarily H$\alpha$) was interpreted by \citet{Heckman1990} in the context of a hot wind. In this case, superheated gas ($T>10^8$~K) in the starburst nucleus adiabatically expands into the ambient medium \citep{Chevalier1985,Strickland2000}. This drives a wind with velocity $v_{\text{wind}} > 1000$~km~s$^{-1}$, with some gas at lower and higher velocities. For a truly adiabatic wind, the asymptotic $v_{\text{wind}} > 3000$~km~s$^{-1}$. The actual velocity will depend on the wind evolution, which remains the subject of active research \citep[see][]{Zhang2018}.
At first, the wind inflates a bubble with a forward shock whose advance speed is faster than the wind fluid, leading to a wind shock. The cooled wind fluid is reheated at the shock to $T \sim 10^6-10^7$~K, so the soft X-rays come from the sheath of shocked wind fluid around the ``free'' wind. Outside of the wind is the shocked CGM, which emits primarily optical and UV lines. Mixing of the shocked wind and shocked ambient medium through Rayleigh-Taylor instabilities also heats the cooler gas and powers the optical and UV emission. 

The morphology of the extended wind in NGC~3079 is in qualitative agreement with this structure, with both the X-rays and FUV forming edge-brightened cones and the X-rays interior to the FUV. We compared the FUV and X-ray morphology to more recent numerical simulations of a hot wind from \citet{Tanner2016}, which follows and is in agreement with earlier work that pioneered fractal gas distributions by \citet{cooper2008}. One such model is shown in Figure~\ref{fig:simulations}and the X-ray and H$\alpha$ (a proxy for FUV) emission look very similar to what we find in NGC~3079 as shown in Figure~\ref{fig:quad_plot}. The authors used the magnetohydrodynamic Athena code \citep{Stone2008} with radiative cooling and photoelectric heating prescriptions. The model seen in Figure~\ref{fig:simulations} shows a projection from 1.5~Myr after the start of the starburst with a SFR of 1.7~$M_{\odot}$~yr$^{-1}$ and an energy injection rate of $7.5\times 10^{41}$~erg~s$^{-1}$. This corresponds to a central mass-loading of $\beta \approx 2$, while the SNe thermalization efficiency is fixed at $\epsilon \equiv 1$. While this model uses parameters similar to M82 and not NGC~3079, the similarity of the structure to what we see in NGC~3079 is encouraging.

We note that the \citet{Tanner2016} simulations include cool, molecular gas, which could, in principle, affect the partitioning of wind energy relative to the \citet{Heckman1990} hot wind model. However, an inspection of the simulations presented here shows that the hot gas contains the bulk of the energy and that the inclusion of molecular gas represents a small perturbation to the energy balance, even though the cold phase contains a substantial fraction of the mass.

Thus, we investigate the wind using the X-ray and FUV data \textit{as if it is produced by a hot wind}. This enables us both to assess whether the data are indeed consistent with a hot wind (i.e., whether interpreting the data in this model leads to sensible results), as well as the impact of this wind on the galaxy and its environs. The values we can infer in this framework are the FUV/X-ray luminosity ratio, the shock velocity (a lower limit on the wind velocity), and the mass-loading factor. We defer a more detailed study of the thermal properties of the wind to a separate paper (Yukita et al., in prep.).

First, we note that the same basic wind structure has also been proposed for AGN-driven winds \citep{Faucher2012}, and NGC~3079 hosts a Compton-thick AGN \citep{Iyomoto2001} with an absorption-corrected $2-10$~keV $L_X \approx 10^{42}$~erg~s$^{-1}$ \citep{LaCaria2019}. Assuming a bolometric correction of 15-25 from \citet{Vasudevan2007}, the true luminosity is a few$\times 10^{43}$~erg~s$^{-1}$, or about 1~SN per year. This is sufficient to drive a galactic wind. The XMM and \Galex\ data alone are insufficient to distinguish between a starburst or AGN origin \citep[see][for a discussion of the wind at smaller radii]{Sebastian2019}. Hereafter, we refer to the hot wind model without regard to the initial power source, except where noted.

\subsection{Luminosity Ratio}

The FUV/X-ray luminosity ratio is consistent with the hot wind model. The 0.3--10~keV X-ray luminosity associated with the wind is $1.1\times 10^{40}$~erg~s$^{-1}$ (Table~\ref{table:xrayprop}), but about half of this comes from the disk region where we cannot isolate the wind FUV emission due to extinction and scattered light. The extraplanar X-ray luminosity is about $6\times 10^{39}$~erg~s$^{-1}$, while the FUV filament luminosity is 23 times higher at $1.4\times 10^{41}$~erg~s$^{-1}$, after correcting for the PSF wings of the galaxy. The disparity intensifies with height. The luminosity ratio should be high in the hot wind model \citep[up to 100;][]{Tanner2016}, as shock-heating and mixing can heat entrained clouds and the CGM to $T \sim 10^5$~K, where the C~{\sc iv} ion fraction is highest, without significantly depleting the thermal energy of the shocked wind. 

This is also consistent with the conclusion from \citet{Borthakur2013} that FUV absorption around starburst galaxies is from collisionally ionized gas. Without spectroscopic information we cannot strongly rule out ionization from the AGN or starburst, but both are disfavored here.

A quasar or more modest AGN outburst could easily ionize the extended wind. For example, \citet{Kreimeyer2013} report giant ionization cones around a quasar, and \citet{Bland-Hawthorn2019} show that activity from Sgr A* likely ionized material in the Magellanic Stream, 75~kpc away from the Galaxy. However, measurements in the prominent superbubble close to the disk \citep{Cecil2001} find an [\ion{O}{3}]/H$\alpha$ ratio of less than 0.4, which is smaller than expected for an AGN \citep{Robitaille2007} and more consistent with a starburst. The AGN is also low luminosity and possibly obscured, so ionization of the extended filaments appear to require a more luminous episode within the past 0.2~Myr, based on the light-travel time to the edge of the nebula. More data are needed to confirm or rule out this possibility.

Despite the soft ionizing spectrum implied by the superbubble line ratios, the starburst is a poor candidate for ionizing the extended nebula. A starburst relies on SNe to power the wind, but the ionizing flux from a given star cluster is highest at very early times before many SNe have exploded. Thus, the starburst would need to continually generate massive, young star clusters and sustain a nuclear wind.

\subsection{Shock Velocities}

In the hot wind paradigm, the wind speed can be estimated from the temperature profile. If the wind rams into adiabatically cooled wind fluid at the wind head or entrained cool gas then the shock will be strong, with $kT \approx \tfrac{3}{16}\mu  v_{\text{wind}}^2$. The temperature profile in NGC~3079 then implies that $v_{\text{wind}}$ decreases from about 570~km~s$^{-1}$ near the disk to 475~km~s$^{-1}$ at 20~kpc above the disk. The outermost points hint that the temperature profile flattens. 

Alternatively, the X-ray emission could be a mixture of shocked wind fluid and shocked ambient CGM. Galaxies like NGC~3079 are believed to have extensive hot CGM with a temperature near or above the virial temperature \citep{Bregman2018}, $kT_{\text{vir}} \approx \tfrac{5}{3} \mu m_p G M_{\text{vir}}/r_{\text{vir}}$, where $M_{\text{vir}} \sim 10^{12} M_{\odot}$ is the mass enclosed in the virial radius and $r_{\text{vir}} \sim 250$~kpc for an $L_*$ galaxy like NGC~3079. This leads to $kT_{\text{vir}} \sim 0.18$~keV, or $2\times 10^6$~K. In this case, the strong shock limit for the Rankine-Hugoniot shock jump conditions may not apply, since even a 1000~km~s$^{-1}$ wind (near the predicted maximum for the adiabatic case) would only shock the CGM at Mach~4. Adopting $0.2$~keV as a rough estimate for the presence of a putative hot halo, the sound speed is $c_s \approx 230$~km~s$^{-1}$ and the temperature profile (Figure~\ref{fig:tprof}) implies Mach numbers of 2.1 to 1.5, or $v_{\text{shock}} \sim 350-480$~km~s$^{-1}$. This is about 20\% smaller than for the strongly shocked case. 

The density profile is independent of the temperature and provides a constraint if we again assume a form for the hot halo. NGC~3079 has a mass similar to that of the Milky Way ($v_{\text{circ}} \approx 208$~km~s$^{-1}$ compared to 240~km~s$^{-1}$ for the Galaxy), so we consider the hot CGM model for the Milky Way from \citet{Miller2015}, which is a modified $\beta$ model with $n(r) = n_0 (r/r_c)^{-3\beta_{\text{CGM}}}$ for $r\gg r_c$, where where $n_0 r_c^{3\beta_{\text{CGM}}} = 0.0135$~cm$^{-3}$~kpc$^{3\beta_{\text{CGM}}}$ and $\beta_{\text{CGM}} \approx 0.5$. The core radius is not independently determined because of confusion towards the Galactic center, so we fixed $\beta_{\text{CGM}} \equiv 0.5$ and determine the $r_c$ and compression factor (Mach number) that best matches the NGC~3079 density profile to the Milky Way density profile. We find a good fit (Figure~\ref{fig:tprof}) for $r_c \sim 6$~kpc and a compression factor of $\approx 2.0$, leading to an average Mach~2.4 shock. 

Of course, both estimates are speculative given the absence of measurements of a hot halo around NGC~3079. Nevertheless, should such a hot halo exist one could still infer the wind velocity from $kT$ as though the X-rays in the wind come from shocked gas. This puts a lower limit on the wind velocity of about 350~km~s$^{-1}$.

Finally, we note that while the \citet{Miller2015} model provides a good match to X-ray emission- and absorption-line data from around the Galaxy, there are several competing models that can qualitatively change the expected wind morphology and shock speed. For example, \citet{Faerman2017} use a combination of absorption and emission data to infer a much more massive halo, while \citet{Gupta2012} and \citet{Nicastro2016} use absorption data sets to describe hot halos that contain upwards of $10^{11} M_{\odot}$. If we perform the same exercise as above for the \citet{Faerman2017} model, which has the advantage of being analytical, we find that it overpredicts the observed density above about 10~kpc while having a similar temperature. If such a massive halo existed around NGC~3079, the high density would likely crush the wind, perhaps turning it into a bubbling plume. However, we note that this does not constitute evidence for or against any Milky Way halo model \citep[for a discussion of those models, see][]{Bregman2018}.

\subsection{Mass Loading and Energetics}

We can estimate the mass-loading factor for gas that reaches high latitudes (here defined as $z>2$~kpc) by comparing the estimated $5\times 10^7 M_{\odot}$ in X-ray emitting gas above this height (Table~\ref{table:xrayprop}) and the size of the wind to the SFR, assuming a starburst origin. The 60~kpc size of the wind and the inferred wind velocity of $\sim$500~km~s$^{-1}$ places a lower bound on the lifetime of 120~Myr (if the X-rays trace shocked gas, the wind cone expands more slowly). Since the average density of hot gas implies a cooling time of hundreds of Myr \citep[assuming Solar metallicity and the collisional equilibrium cooling functions from][]{Gnat2007}, the X-rays should trace the energy deposited by the shocked wind over its history.

The 120~Myr limit implies an outflow rate to high latitudes of $\dot{M} < 0.4 M_{\odot}$~yr$^{-1}$. If we further assume that the current SFR$\approx$2.6~$M_{\odot}$~yr$^{-1}$ \citep{Yamagishi2010} is the average over that period and that the central mass-loading factor is $\beta = 1$, then the high-latitude $\beta_{\text{hl}} < 0.2$. Note that even if we assume a younger wind (60~Myr at a  terminal velocity of 1000~km~s$^{-1}$), $\beta_{\text{hl}} < 0.4$. However, $\beta_{\text{hl}}$ will be increased by the mass in the swept-up shell that is accelerated outwards. We do not know the photoionization fraction, so we have a poor constraint on the mass represented by the FUV emission. If all FUV-emitting material is shock-heated and accelerated outwards, then it is possible for $\beta_{\text{hl}}$ to approach unity. 

The mass and inferred shock speed lead to a kinetic energy of about $1\times 10^{56}$~erg. For a 120~Myr lifetime, the kinetic luminosity of the high-latitude wind is less than $2.7\times 10^{40}$~erg~s$^{-1}$. If we further assume that 1\% of the stellar mass becomes SNe, that each SN contributes $10^{51}$~erg, and that the thermalization efficiency is unity, then the average kinetic luminosity of the starburst is $8\times 10^{41}$~erg~s$^{-1}$. This is consistent with a Starburst99 \citep{Leitherer1999} population synthesis model for the same input SFR at solar metallicity. The average wind kinetic luminosity is only 3.4\% of this value. We have not considered the UV contribution, but it will be small. Strongly shocked gas transforms roughly half the kinetic energy into thermal energy, so even with zero photoionization and ten times more mass in warm gas the contribution will be smaller than from the X-ray emitting gas.

The 120~Myr estimate from assuming a terminal wind velocity of 500~km~s$^{-1}$ is in tension with other estimates for the starburst age. \citet{Yamagishi2010} found that the gas-to-dust ratio is rather high and concluded that NGC~3079 is early in its starburst phase. Meanwhile, \citet{Konami2012} defined a region with $\alpha$-enhanced abundances (an annulus centered at 4.5~kpc from the nucleus) and used the hydrodynamic model of \citet{Tomisaka1993} to estimate the expansion velocity, and thus the age. They arrived at a velocity of about 450~km~s$^{-1}$, which is consistent with our shock-inferred velocity but leads to a starburst age of 10~Myr. This is consistent with the age of the well known superbubble \citet{Cecil2001}.

There are a few ways to reconcile the two ages. The 120~Myr value assumes that the overall wind terminal velocity is 500~km~s$^{-1}$, which is substantially below the 3000~km~s$^{-1}$ adiabatic solution or 1000~km~s$^{-1}$ from \citet{Strickland2000}. Gas that remains this fast to large radii will be hard to see and must not strongly interact with the surrounding medium, or it would dissipate its kinetic energy. Thus, there may be a very fast spine to the wind with shocks at the edges from lower velocity components. However, for a maximal expansion velocity of 3000~km~s$^{-1}$, the starburst or AGN outburst would still need to be at least 20~Myr old.

Alternatively, the extended structure may be a relic from a prior outburst. If it were a jetted AGN outburst, we would expect to see low-frequency radio synchrotron emission from aging cosmic rays. The 326~MHz radio continuum maps in \citet{Irwin2003} reveal a large radio halo but no structures close to the scale of the wind described here. Instead, the well known radio lobes occur on much smaller scales near the disk, and larger extensions that may indicate a prior outburst are about 10~kpc in size.

Finally, the shocked wind interpretation that we applied above may be wrong. A different source for the X-rays and FUV would lead to different constraints on the wind speed. Reconciling the young starburst with the large filaments is a puzzle that remains to be solved.

\subsection{A Hot Wind?}

The FUV and X-ray morphology, FUV/X-ray luminosity ratio, and small $\beta$ at high latitudes are consistent with a fast, non-radiative wind powered by superheated gas. On the other hand, the inferred velocity and kinetic energy are smaller than expected for an adiabatic wind. In such a wind, $v_{\text{wind}} \sim 1000 \epsilon^{1/2} \beta^{-1/2}$~km~s$^{-1}$, where $\epsilon \in [0,1]$ is the SNe thermalization efficiency and $\beta$ is the mass-loading factor. To reconcile this velocity with the $\sim$500~km~s$^{-1}$ implied by the X-ray temperature requires $\epsilon \approx 0.25$ or a large $\beta = 4$. For a powerful wind, a large $\beta$ is more likely than a small $\epsilon$, and the high latitude $\beta_{\text{hl}} < 0.2$ would mean that very little of the mass makes it to high latitudes and the wind ceases to be adiabatic at lower latitudes. This scenario is supported by the density profile, which is inconsistent with the $R^{-2}$ profile (or any power-law model) expected from the adiabatic model. Since the shocked wind is produced by the wind overrunning itself, the density profile from the X-rays should be (to first order) a compressed version of the underlying profile. 

It would not be surprising for a wind that begins with adiabatic free expansion to leave the adiabatic regime after several kpc, so our main concern is whether inferring $v_{\text{shock}}$ from $kT$ is still valid in this case, i.e., whether the high latitude wind is still fast and non-radiative. \citet{Strickland2000} found that an initially adiabatic wind will not remain so but that X-rays still predominantly come from shocked gas, with some reasonable parameter sets (based on M82) yielding an average wind speed of $v_{\text{wind}} \sim 500$~km~s$^{-1}$. This suggests that $kT$ should roughly map to $v_{\text{wind}}$ even when the adiabatic approximation breaks down. 

However, there are two more problems for the hot wind. First, the density profile suggests that the FUV and X-ray emission are coming from shocked CGM rather than shocked wind, or that the shocked wind contributes little mass and mixes efficiently with the CGM. Secondly, the kinetic energy implied by the velocity and density places an upper bound on the kinetic luminosity of $E<2.7\times 10^{40}$~erg~s$^{-1}$, which is less than 3.4\% of the total starburst kinetic luminosity. In contrast, hot wind models hold that most of the wind energy is in hot gas. 

In summary, the morphology and luminosities are consistent with a fast, non-radiative wind that shock heats the surroundings, but that carries only a small fraction of the starburst or AGN energy, likely due to carrying a small amount of mass. In this case, the remainder of the energy would accelerate cooler gas at lower latitudes in an action like a galactic fountain. Alternatively, if a collimated wind propagates far beyond the 60~kpc limit to the FUV emission, then the forward shock may effectively cease to exist in the tenuous CGM and the X-rays could represent only a small fraction of the energy carried by the wind. This would be the case if there is a 3000~km~s$^{-1}$ component that persists to large radii.

\subsection{Impact}

Regardless of the nature of the wind, we can draw several conclusions about its impact on the galaxy. First, the $5\times 10^7 M_{\odot}$ in hot, high-latitude gas implies that the wind is not effective at removing mass from the galaxy. If the starburst is self-limiting due to feedback, the heating and recycling occurs at radii of at most a few kpc. The masses and limits from above lead to a rate of $<$1~$M_{\odot}$~yr$^{-1}$ for removing gas to $\gtrsim$2~kpc \citep[deep radio observations place limits on the neutral hydrogen in the wind;][]{Shafi2015}. The SFR is $\approx 2.6 M_{\odot}$~yr$^{-1}$, so much more gas will be locked in stars than completely removed from the galaxy by the end of the starburst. 

Secondly, the wind may indeed heat the intergalactic medium, assuming that $kT$ maps to $v_{\text{shock}}$. At 500~km~s$^{-1}$ between 1-30~kpc from the disk, the wind very likely exceeds the escape velocity. A conservative estimate for the escape velocity within the disk, $v_{\text{esc}} \sim 3 v_{\text{circ}} = 625$~km~s$^{-1}$, implies a lower $v_{\text{esc}} < 550$~km~s$^{-1}$ at a height of 1~kpc for an exponential disk model. Most of the X-rays from the central bin come from at least this height (in large part due to absorption in the edge-on disk). At larger heights, a disk model may not be appropriate. NGC~3079 has a similar stellar mass and $v_{\text{circ}}$ to the Milky Way, so if we adopt the RAVE Milky Way mass \citep{Piffl2014} and a NFW profile, the escape velocity at 20~kpc is $250-300$~km~s$^{-1}$. Hence, the $<$1~$M_{\odot}$~yr$^{-1}$ expelled to $>$2~kpc from the disk is also the limit on gas unbound from the galaxy. 

Thirdly, the wind has a strong impact on the CGM. The density profile is incompatible with the adiabatic approximation ($n(r) \propto R^{-2}$) and this suggests that many of the X-rays come from shock-compressed CGM or entrained clouds rather than strictly shocked wind.  The 10~kpc scale height of the best-fit exponential profile suggests that it is mostly CGM, since most cold ISM is not expelled to high latitudes. 10~kpc is rather large for a relaxed, disk-like atmosphere around a galaxy, but the extended CGM may follow a spherical distribution like the $\beta$ model, in which $n(r) \propto (1+(r/r_c)^2)^{-3\beta_{\text{CGM}}/2}$. This is a good fit to the data for $r_c \sim 6$~kpc when $\beta_{\text{CGM}} \equiv 0.5$ (Figure~\ref{fig:tprof}) regardless of whether the working surface is warm or hot CGM. 

If the working surface is warm CGM then only a small fraction of the mass is heated, but all of the CGM traced by the FUV is compressed, which may trigger condensation and infall of clouds. On the other hand, if the working surface is hot CGM then the kinetic luminosity of $\dot{E}<2.7\times 10^{40}$~erg~s$^{-1}$ implied by the hot gas is likely substantially higher than the few$\times 10^{39}$~erg~s$^{-1}$ radiative luminosity of a normal hot halo in an $L_*$ galaxy \citep{JiangtaoLi2013} and can prevent cooling. Even if the leading edge of the wind expands 10~times slower than its internal speed, the high-latitude mechanical luminosity is more than enough to balance radiative cooling. The other major impact on the hot CGM is through displacement. Although the opening angle is only about 30$^{\circ}$, the hot mass in the wind cone is similar to that expected from the \citet{Miller2015} Milky Way model within a radius of about 8~kpc. Thus, as the wind evolves and eventually dissipates the hot and expanding cone will continue to stir up the CGM for a long time.

The wind also likely affects the companion galaxy, NGC~3073. \citet{Shafi2015} argued that an ambient hot density of $1.3\times 10^{-2}$~cm$^{-2}$ is needed to explain the cometary \ion{H}{1} tail of NGC~3073 through ram-pressure stripping due to infall. The hot density profile shows that the hot density around NGC~3073 (beyond 25~kpc from NGC~3079) is at least an order of magnitude too small. In contrast, the ram pressure from $v_{\text{wind}} \gtrsim 500$~km~s$^{-1}$ and a density of $\sim 5\times 10^{-4}$~cm$^{-3}$ is likely, but barely, sufficient to produce the tail based on the arguments in \citet{Irwin1987}. If NGC~3073 is in the middle of the wind, then $v_{\text{wind}}$ could be substantially higher, which would strengthen the case for wind stripping.

\section{Summary and Conclusions}
\label{section.summary}

We report the discovery of FUV emission around NGC~3079 at least to 60~kpc from the nucleus, with X-rays detected to at least 30~kpc. The FUV and X-ray emission is biconical and edge-brightened, with the X-rays interior to the FUV. We rule out dust scattering as the source of the FUV light, which makes line emission the most likely candidate. Meanwhile, the X-ray spectrum is consistent with line-dominated thermal emission from a plasma near collisional ionization equilibrium. We measured the temperature and density of the hot gas. Both decline with height, with a possible flattening in the temperature beyond 20~kpc at $kT \approx 0.27$~keV. The total X-ray emitting mass is about $5\times 10^7 M_{\odot}$. 

The extended FUV and X-ray emission connects smoothly to emission already well known at lower latitudes, and is part of a galactic superwind. The morphology of the FUV and X-ray emission suggest a hot, non-radiative wind in which the emission is produced by shock heating (rather than radiative cooling of upstream wind fluid). Assuming this to be the case, the X-ray temperatures imply wind velocities of $\sim$500~km~s$^{-1}$, which is sufficient to escape the galaxy. However, the density profile and low mass and kinetic energy in hot gas are inconsistent with a hot wind model, so the nature of the wind must be further explored (Yukita et al., in prep.). 

The wind carries little mass and less kinetic energy than expected for a hot wind, but will nonetheless significantly heat the CGM and perhaps the IGM. It remains an open question whether winds, or how much of them, can escape the galaxy's potential. If the extended emission in NGC~3079 traces shock-heated gas, then the wind is able to maintain high velocities at least to 20\% of the virial radius, and will encounter less resistance the farther it goes. 

The extended wind reported here is one of just a handful of highly extended winds, with others including NGC~6240 \citep{Yoshida2016} and Makani \citep{Rupke2019}. NGC~3079 is by far the closest, and presents a good target for deep H$\alpha$ mapping. The serendipitous discovery of the extended filaments in the deep \Galex\ data suggests searching for extended emission around other nearby, well developed starbursts, such as NGC~253 or M82. These galaxies are even closer, so a very wide field is required. 

The other work that is required is modeling of the evolution of winds in realistic CGM environments and over long times. This is difficult because the high resolution required to resolve instabilities and mixing makes a large box computationally expensive. However, adaptive mesh refinement and similar techniques for concentrating resolution where it is needed may make these models feasible. 

\acknowledgments

We thank the anonymous referee for a careful and helpful report which improved this paper. M.~Y. gratefully acknowledges support through NASA grant 80NSSC18K0609. This research has made use of the NASA/IPAC Extragalactic Database (NED) which is operated by the Jet Propulsion Laboratory, California Institute of Technology, under contract with the National Aeronautics and Space Administration. 

\bibliographystyle{aasjournal}
%\bibliography{references}

\begin{thebibliography}{}
\expandafter\ifx\csname natexlab\endcsname\relax\def\natexlab#1{#1}\fi
\providecommand{\url}[1]{\href{#1}{#1}}
\providecommand{\dodoi}[1]{doi:~\href{http://doi.org/#1}{\nolinkurl{#1}}}
\providecommand{\doeprint}[1]{\href{http://ascl.net/#1}{\nolinkurl{http://ascl.net/#1}}}
\providecommand{\doarXiv}[1]{\href{https://arxiv.org/abs/#1}{\nolinkurl{https://arxiv.org/abs/#1}}}

\bibitem[{{Anders} \& {Grevesse}(1989)}]{Anders1989}
{Anders}, E., \& {Grevesse}, N. 1989, \gca, 53, 197,
  \dodoi{10.1016/0016-7037(89)90286-X}

\bibitem[{{Anderson} \& {Bregman}(2014)}]{Anderson2014}
{Anderson}, M.~E., \& {Bregman}, J.~N. 2014, \apj, 785, 67,
  \dodoi{10.1088/0004-637X/785/1/67}

\bibitem[{{Asplund} {et~al.}(2009){Asplund}, {Grevesse}, {Sauval}, \&
  {Scott}}]{Asplund2009}
{Asplund}, M., {Grevesse}, N., {Sauval}, A.~J., \& {Scott}, P. 2009, \araa, 47,
  481, \dodoi{10.1146/annurev.astro.46.060407.145222}

\bibitem[{{Bland-Hawthorn} {et~al.}(2019){Bland-Hawthorn}, {Maloney},
  {Sutherland}, {Groves}, {Guglielmo}, {Li}, {Curzons}, {Cecil}, \&
  {Fox}}]{Bland-Hawthorn2019}
{Bland-Hawthorn}, J., {Maloney}, P.~R., {Sutherland}, R., {et~al.} 2019, \apj,
  886, 45, \dodoi{10.3847/1538-4357/ab44c8}

\bibitem[{{Bordoloi} {et~al.}(2014){Bordoloi}, {Lilly}, {Kacprzak}, \&
  {Churchill}}]{Bordoloi2014}
{Bordoloi}, R., {Lilly}, S.~J., {Kacprzak}, G.~G., \& {Churchill}, C.~W. 2014,
  \apj, 784, 108, \dodoi{10.1088/0004-637X/784/2/108}

\bibitem[{{Borthakur} {et~al.}(2013){Borthakur}, {Heckman}, {Strickland},
  {Wild}, \& {Schiminovich}}]{Borthakur2013}
{Borthakur}, S., {Heckman}, T., {Strickland}, D., {Wild}, V., \&
  {Schiminovich}, D. 2013, \apj, 768, 18, \dodoi{10.1088/0004-637X/768/1/18}

\bibitem[{{Bregman} {et~al.}(2018){Bregman}, {Anderson}, {Miller},
  {Hodges-Kluck}, {Dai}, {Li}, {Li}, \& {Qu}}]{Bregman2018}
{Bregman}, J.~N., {Anderson}, M.~E., {Miller}, M.~J., {et~al.} 2018, \apj, 862,
  3, \dodoi{10.3847/1538-4357/aacafe}

\bibitem[{{Cecil} {et~al.}(2002){Cecil}, {Bland-Hawthorn}, \&
  {Veilleux}}]{Cecil2002}
{Cecil}, G., {Bland-Hawthorn}, J., \& {Veilleux}, S. 2002, \apj, 576, 745,
  \dodoi{10.1086/341861}

\bibitem[{{Cecil} {et~al.}(2001){Cecil}, {Bland-Hawthorn}, {Veilleux}, \&
  {Filippenko}}]{Cecil2001}
{Cecil}, G., {Bland-Hawthorn}, J., {Veilleux}, S., \& {Filippenko}, A.~V. 2001,
  \apj, 555, 338, \dodoi{10.1086/321481}

\bibitem[{{Chevalier} \& {Clegg}(1985)}]{Chevalier1985}
{Chevalier}, R.~A., \& {Clegg}, A.~W. 1985, \nat, 317, 44,
  \dodoi{10.1038/317044a0}

\bibitem[{{Cooper} {et~al.}(2008){Cooper}, {Bicknell}, {Sutherland }, \&
  {Bland-Hawthorn}}]{cooper2008}
{Cooper}, J.~L., {Bicknell}, G.~V., {Sutherland }, R.~S., \& {Bland-Hawthorn},
  J. 2008, \apj, 674, 157, \dodoi{10.1086/524918}

\bibitem[{{Duval} {et~al.}(2016){Duval}, {{\"O}stlin}, {Hayes}, {Zackrisson},
  {Verhamme}, {Orlitova}, {Adamo}, {Guaita}, {Melinder}, {Cannon}, {Laursen},
  {Rivera-Thorsen}, {Herenz}, {Gruyters}, {Mas-Hesse}, {Kunth}, {Sandberg},
  {Schaerer}, \& {M{\r{a}}nsson}}]{Duval2016}
{Duval}, F., {{\"O}stlin}, G., {Hayes}, M., {et~al.} 2016, \aap, 587, A77,
  \dodoi{10.1051/0004-6361/201526876}

\bibitem[{{Fabbiano} {et~al.}(1992){Fabbiano}, {Kim}, \&
  {Trinchieri}}]{Fabbiano1992}
{Fabbiano}, G., {Kim}, D.~W., \& {Trinchieri}, G. 1992, \apjs, 80, 531,
  \dodoi{10.1086/191675}

\bibitem[{{Faerman} {et~al.}(2017){Faerman}, {Sternberg}, \&
  {McKee}}]{Faerman2017}
{Faerman}, Y., {Sternberg}, A., \& {McKee}, C.~F. 2017, \apj, 835, 52,
  \dodoi{10.3847/1538-4357/835/1/52}

\bibitem[{{Faucher-Gigu{\`e}re} \& {Quataert}(2012)}]{Faucher2012}
{Faucher-Gigu{\`e}re}, C.-A., \& {Quataert}, E. 2012, \mnras, 425, 605,
  \dodoi{10.1111/j.1365-2966.2012.21512.x}

\bibitem[{{Ferland} {et~al.}(2017){Ferland}, {Chatzikos}, {Guzm{\'a}n},
  {Lykins}, {van Hoof}, {Williams}, {Abel}, {Badnell}, {Keenan}, {Porter}, \&
  {Stancil}}]{Ferland2017}
{Ferland}, G.~J., {Chatzikos}, M., {Guzm{\'a}n}, F., {et~al.} 2017, \rmxaa, 53,
  385.
\newblock \doarXiv{1705.10877}

\bibitem[{{Finley} {et~al.}(2017){Finley}, {Bouch{\'e}}, {Contini}, {Epinat},
  {Bacon}, {Brinchmann}, {Cantalupo}, {Erroz-Ferrer}, {Marino}, {Maseda},
  {Richard}, {Schroetter}, {Verhamme}, {Weilbacher}, {Wendt}, \&
  {Wisotzki}}]{Finley2017}
{Finley}, H., {Bouch{\'e}}, N., {Contini}, T., {et~al.} 2017, \aap, 605, A118,
  \dodoi{10.1051/0004-6361/201730428}

\bibitem[{{Gnat} \& {Sternberg}(2007)}]{Gnat2007}
{Gnat}, O., \& {Sternberg}, A. 2007, \apjs, 168, 213, \dodoi{10.1086/509786}

\bibitem[{{Gupta} {et~al.}(2012){Gupta}, {Mathur}, {Krongold}, {Nicastro}, \&
  {Galeazzi}}]{Gupta2012}
{Gupta}, A., {Mathur}, S., {Krongold}, Y., {Nicastro}, F., \& {Galeazzi}, M.
  2012, \apjl, 756, L8, \dodoi{10.1088/2041-8205/756/1/L8}

\bibitem[{{Heckman} {et~al.}(2017){Heckman}, {Borthakur}, {Wild},
  {Schiminovich}, \& {Bordoloi}}]{Heckman2017b}
{Heckman}, T., {Borthakur}, S., {Wild}, V., {Schiminovich}, D., \& {Bordoloi},
  R. 2017, \apj, 846, 151, \dodoi{10.3847/1538-4357/aa80dc}

\bibitem[{{Heckman} {et~al.}(1990){Heckman}, {Armus}, \& {Miley}}]{Heckman1990}
{Heckman}, T.~M., {Armus}, L., \& {Miley}, G.~K. 1990, \apjs, 74, 833,
  \dodoi{10.1086/191522}

\bibitem[{{Heckman} \& {Thompson}(2017)}]{Heckman2017}
{Heckman}, T.~M., \& {Thompson}, T.~A. 2017, {Galactic Winds and the Role
  Played by Massive Stars}, ed. A.~W. {Alsabti} \& P.~{Murdin}, 2431,
  \dodoi{10.1007/978-3-319-21846-5_23}

\bibitem[{{Hodges-Kluck} {et~al.}(2016){Hodges-Kluck}, {Cafmeyer}, \&
  {Bregman}}]{Hodges-Kluck2016}
{Hodges-Kluck}, E., {Cafmeyer}, J., \& {Bregman}, J.~N. 2016, \apj, 833, 58,
  \dodoi{10.3847/1538-4357/833/1/58}

\bibitem[{{Irwin} \& {Saikia}(2003)}]{Irwin2003}
{Irwin}, J.~A., \& {Saikia}, D.~J. 2003, \mnras, 346, 977,
  \dodoi{10.1111/j.1365-2966.2003.07146.x}

\bibitem[{{Irwin} {et~al.}(1987){Irwin}, {Seaquist}, {Taylor}, \&
  {Duric}}]{Irwin1987}
{Irwin}, J.~A., {Seaquist}, E.~R., {Taylor}, A.~R., \& {Duric}, N. 1987, \apjl,
  313, L91, \dodoi{10.1086/184837}

\bibitem[{{Iyomoto} {et~al.}(2001){Iyomoto}, {Fukazawa}, {Nakai}, \&
  {Ishihara}}]{Iyomoto2001}
{Iyomoto}, N., {Fukazawa}, Y., {Nakai}, N., \& {Ishihara}, Y. 2001, \apjl, 561,
  L69, \dodoi{10.1086/324056}

\bibitem[{{Jaskot} \& {Ravindranath}(2016)}]{Jaskot2016}
{Jaskot}, A.~E., \& {Ravindranath}, S. 2016, \apj, 833, 136,
  \dodoi{10.3847/1538-4357/833/2/136}

\bibitem[{{Kacprzak} {et~al.}(2012){Kacprzak}, {Churchill}, \&
  {Nielsen}}]{Kacprzak2012}
{Kacprzak}, G.~G., {Churchill}, C.~W., \& {Nielsen}, N.~M. 2012, \apjl, 760,
  L7, \dodoi{10.1088/2041-8205/760/1/L7}

\bibitem[{{Konami} {et~al.}(2012){Konami}, {Matsushita}, {Gandhi}, \&
  {Tamagawa}}]{Konami2012}
{Konami}, S., {Matsushita}, K., {Gandhi}, P., \& {Tamagawa}, T. 2012, \pasj,
  64, 117, \dodoi{10.1093/pasj/64.5.117}

\bibitem[{{Kreimeyer} \& {Veilleux}(2013)}]{Kreimeyer2013}
{Kreimeyer}, K., \& {Veilleux}, S. 2013, \apjl, 772, L11,
  \dodoi{10.1088/2041-8205/772/1/L11}

\bibitem[{{La Caria} {et~al.}(2019){La Caria}, {Vignali}, {Lanzuisi},
  {Gruppioni}, \& {Pozzi}}]{LaCaria2019}
{La Caria}, M.~M., {Vignali}, C., {Lanzuisi}, G., {Gruppioni}, C., \& {Pozzi},
  F. 2019, \mnras, 487, 1662, \dodoi{10.1093/mnras/stz1381}

\bibitem[{{Leitherer} {et~al.}(1999){Leitherer}, {Schaerer}, {Goldader},
  {Delgado}, {Robert}, {Kune}, {de Mello}, {Devost}, \&
  {Heckman}}]{Leitherer1999}
{Leitherer}, C., {Schaerer}, D., {Goldader}, J.~D., {et~al.} 1999, \apjs, 123,
  3, \dodoi{10.1086/313233}

\bibitem[{{Li} {et~al.}(2019){Li}, {Hodges-Kluck}, {Stein}, {Bregman}, {Irwin},
  \& {Dettmar}}]{JiangtaoLi2019}
{Li}, J.-T., {Hodges-Kluck}, E., {Stein}, Y., {et~al.} 2019, \apj, 873, 27,
  \dodoi{10.3847/1538-4357/ab010a}

\bibitem[{{Li} \& {Wang}(2013)}]{JiangtaoLi2013}
{Li}, J.-T., \& {Wang}, Q.~D. 2013, \mnras, 435, 3071,
  \dodoi{10.1093/mnras/stt1501}

\bibitem[{{Miller} \& {Bregman}(2015)}]{Miller2015}
{Miller}, M.~J., \& {Bregman}, J.~N. 2015, \apj, 800, 14,
  \dodoi{10.1088/0004-637X/800/1/14}

\bibitem[{{Miville-Desch{\^e}nes} \& {Lagache}(2005)}]{Miville-Deschenes2005}
{Miville-Desch{\^e}nes}, M.-A., \& {Lagache}, G. 2005, \apjs, 157, 302,
  \dodoi{10.1086/427938}

\bibitem[{{Nicastro} {et~al.}(2016){Nicastro}, {Senatore}, {Krongold},
  {Mathur}, \& {Elvis}}]{Nicastro2016}
{Nicastro}, F., {Senatore}, F., {Krongold}, Y., {Mathur}, S., \& {Elvis}, M.
  2016, \apjl, 828, L12, \dodoi{10.3847/2041-8205/828/1/L12}

\bibitem[{{Nomoto} {et~al.}(2006){Nomoto}, {Tominaga}, {Umeda}, {Kobayashi}, \&
  {Maeda}}]{Nomoto2006}
{Nomoto}, K., {Tominaga}, N., {Umeda}, H., {Kobayashi}, C., \& {Maeda}, K.
  2006, \nphysa, 777, 424, \dodoi{10.1016/j.nuclphysa.2006.05.008}

\bibitem[{{Piffl} {et~al.}(2014){Piffl}, {Scannapieco}, {Binney}, {Steinmetz},
  {Scholz}, {Williams}, {de Jong}, {Kordopatis}, {Matijevi{\v c}},
  {Bienaym{\'e}}, {Bland-Hawthorn}, {Boeche}, {Freeman}, {Gibson}, {Gilmore},
  {Grebel}, {Helmi}, {Munari}, {Navarro}, {Parker}, {Reid}, {Seabroke},
  {Watson}, {Wyse}, \& {Zwitter}}]{Piffl2014}
{Piffl}, T., {Scannapieco}, C., {Binney}, J., {et~al.} 2014, \aap, 562, A91,
  \dodoi{10.1051/0004-6361/201322531}

\bibitem[{{Robitaille} {et~al.}(2007){Robitaille}, {Rossa}, {Bomans}, \& {van
  der Marel}}]{Robitaille2007}
{Robitaille}, T.~P., {Rossa}, J., {Bomans}, D.~J., \& {van der Marel}, R.~P.
  2007, \aap, 464, 541, \dodoi{10.1051/0004-6361:20065454}

\bibitem[{{Rupke}(2018)}]{Rupke2018}
{Rupke}, D. 2018, Galaxies, 6, 138, \dodoi{10.3390/galaxies6040138}

\bibitem[{{Rupke} {et~al.}(2019){Rupke}, {Coil}, {Geach}, {Tremonti},
  {Diamond-Stanic}, {George}, {Hickox}, {Kepley}, {Leung}, {Moustakas},
  {Rudnick}, \& {Sell}}]{Rupke2019}
{Rupke}, D. S.~N., {Coil}, A., {Geach}, J.~E., {et~al.} 2019, \nat, 574, 643,
  \dodoi{10.1038/s41586-019-1686-1}

\bibitem[{{Sebastian} {et~al.}(2019){Sebastian}, {Kharb}, {O'Dea}, {Colbert},
  \& {Baum}}]{Sebastian2019}
{Sebastian}, B., {Kharb}, P., {O'Dea}, C.~P., {Colbert}, E.~J.~M., \& {Baum},
  S.~A. 2019, \apj, 883, 189, \dodoi{10.3847/1538-4357/ab371a}

\bibitem[{{Shafi} {et~al.}(2015){Shafi}, {Oosterloo}, {Morganti},
  {Colafrancesco}, \& {Booth}}]{Shafi2015}
{Shafi}, N., {Oosterloo}, T.~A., {Morganti}, R., {Colafrancesco}, S., \&
  {Booth}, R. 2015, \mnras, 454, 1404, \dodoi{10.1093/mnras/stv2034}

\bibitem[{{Snowden} {et~al.}(2004){Snowden}, {Collier}, \&
  {Kuntz}}]{Snowden2004}
{Snowden}, S.~L., {Collier}, M.~R., \& {Kuntz}, K.~D. 2004, \apj, 610, 1182,
  \dodoi{10.1086/421841}

\bibitem[{{Sofue} {et~al.}(2001){Sofue}, {Koda}, {Kohno}, {Okumura}, {Honma},
  {Kawamura}, \& {Irwin}}]{Sofue2001}
{Sofue}, Y., {Koda}, J., {Kohno}, K., {et~al.} 2001, \apjl, 547, L115,
  \dodoi{10.1086/318907}

\bibitem[{{Springob} {et~al.}(2009){Springob}, {Masters}, {Haynes},
  {Giovanelli}, \& {Marinoni}}]{Springob2009}
{Springob}, C.~M., {Masters}, K.~L., {Haynes}, M.~P., {Giovanelli}, R., \&
  {Marinoni}, C. 2009, \apjs, 182, 474, \dodoi{10.1088/0067-0049/182/1/474}

\bibitem[{{Stone} {et~al.}(2008){Stone}, {Gardiner}, {Teuben}, {Hawley}, \&
  {Simon}}]{Stone2008}
{Stone}, J.~M., {Gardiner}, T.~A., {Teuben}, P., {Hawley}, J.~F., \& {Simon},
  J.~B. 2008, \apjs, 178, 137, \dodoi{10.1086/588755}

\bibitem[{{Strickland} {et~al.}(2004){Strickland}, {Heckman}, {Colbert},
  {Hoopes}, \& {Weaver}}]{Strickland2004}
{Strickland}, D.~K., {Heckman}, T.~M., {Colbert}, E.~J.~M., {Hoopes}, C.~G., \&
  {Weaver}, K.~A. 2004, \apjs, 151, 193, \dodoi{10.1086/382214}

\bibitem[{{Strickland} \& {Stevens}(2000)}]{Strickland2000}
{Strickland}, D.~K., \& {Stevens}, I.~R. 2000, \mnras, 314, 511,
  \dodoi{10.1046/j.1365-8711.2000.03391.x}

\bibitem[{{Tanner} {et~al.}(2016){Tanner}, {Cecil}, \& {Heitsch}}]{Tanner2016}
{Tanner}, R., {Cecil}, G., \& {Heitsch}, F. 2016, \apj, 821, 7,
  \dodoi{10.3847/0004-637X/821/1/7}

\bibitem[{{Thompson} {et~al.}(2016){Thompson}, {Quataert}, {Zhang}, \&
  {Weinberg}}]{Thompson2016}
{Thompson}, T.~A., {Quataert}, E., {Zhang}, D., \& {Weinberg}, D.~H. 2016,
  \mnras, 455, 1830, \dodoi{10.1093/mnras/stv2428}

\bibitem[{{Tomisaka} \& {Bregman}(1993)}]{Tomisaka1993}
{Tomisaka}, K., \& {Bregman}, J.~N. 1993, \pasj, 45, 513.
\newblock \doarXiv{astro-ph/9209002}

\bibitem[{{Vasudevan} \& {Fabian}(2007)}]{Vasudevan2007}
{Vasudevan}, R.~V., \& {Fabian}, A.~C. 2007, \mnras, 381, 1235,
  \dodoi{10.1111/j.1365-2966.2007.12328.x}

\bibitem[{{Veilleux} {et~al.}(2005){Veilleux}, {Cecil}, \&
  {Bland-Hawthorn}}]{Veilleux2005}
{Veilleux}, S., {Cecil}, G., \& {Bland-Hawthorn}, J. 2005, \araa, 43, 769,
  \dodoi{10.1146/annurev.astro.43.072103.150610}

\bibitem[{{Yamagishi} {et~al.}(2010){Yamagishi}, {Kaneda}, {Ishihara},
  {Komugi}, {Suzuki}, \& {Onaka}}]{Yamagishi2010}
{Yamagishi}, M., {Kaneda}, H., {Ishihara}, D., {et~al.} 2010, \pasj, 62, 1085,
  \dodoi{10.1093/pasj/62.4.1085}

\bibitem[{{Yoshida} {et~al.}(2016){Yoshida}, {Yagi}, {Ohyama}, {Komiyama},
  {Kashikawa}, {Tanaka}, \& {Okamura}}]{Yoshida2016}
{Yoshida}, M., {Yagi}, M., {Ohyama}, Y., {et~al.} 2016, \apj, 820, 48,
  \dodoi{10.3847/0004-637X/820/1/48}

\bibitem[{{Zhang}(2018)}]{Zhang2018}
{Zhang}, D. 2018, Galaxies, 6, 114, \dodoi{10.3390/galaxies6040114}

\end{thebibliography}

\end{document}